\documentclass[11pt]{article}  
\usepackage[utf8]{inputenc}
\usepackage{amsmath,amssymb}
\usepackage{graphicx}
\usepackage{color}
\usepackage[authoryear,round]{natbib}
\usepackage{fullpage}

\title{Robust identification of local adaptation from allele frequencies}
\author{
Torsten G\"unther$^1$ and Graham Coop$^{2}$ \\
$^1$ Institute of Plant Breeding, Seed Science and Population Genetics,\\ University of Hohenheim, Stuttgart, Germany.\\
$^2$ Department of Evolution and Ecology \& Center for Population Biology,\\ University of California, Davis, USA.\\
\small To whom correspondence should be addressed: \texttt{torsten.guenther@uni-hohenheim.de, gmcoop@ucdavis.edu}\\
}

\date{}

\begin{document}

\maketitle

\begin{abstract}
Comparing allele frequencies among populations that differ in environment has long been a tool for detecting loci involved in local adaptation. However, such analyses are complicated by an imperfect knowledge of population allele frequencies and neutral correlations of allele frequencies among populations due to shared population history and gene flow. Here we develop a set of methods to robustly test for unusual allele frequency patterns, and correlations between environmental variables and allele frequencies while accounting for these complications based on a Bayesian model previously implemented in the software Bayenv. Using this model, we calculate a set of `standardized allele frequencies' that allows investigators to apply tests of their choice to multiple populations, while accounting for sampling and covariance due to population history. We illustrate this first by showing that these standardized frequencies can be used to calculate powerful tests to detect non-parametric correlations with environmental variables, which are also less prone to spurious results due to outlier populations. We then demonstrate how these standardized allele frequencies can be used to construct a test to detect SNPs that deviate strongly from neutral population structure. This test is conceptually related to $F_{ST}$ but should be more powerful as we account for population history. We also extend the model to next-generation sequencing of population pools, which is a cost-efficient way to estimate population allele frequencies, but it implies an additional level of sampling noise. The utility of these methods is demonstrated in simulations and by re-analyzing human SNP data from the HGDP populations. An implementation of our method is available from \texttt{http://gcbias.org}

\end{abstract}

\section{Introduction}
The phenotypes of individuals within a species often vary clinally along environmental gradients \citep{Huxley1939}. Such phenotypic clines have long been central to adaptive arguments in evolutionary biology, with many diverse examples including skin pigmentation in humans \citep{Jablonski2004}, body size and temperature tolerance in \textit{Drosophila} \citep{Hoffmann2007}, and flowering time in plants \citep{Stinchcombe2004}, which all vary clinally with latitude. Unsurprisingly, comparisons of allele frequencies between populations that differ in environment were among the earliest population genetic tests for selection \citep{Cavalli-Sforza1966,Lewontin1973}, and have continued to be central to population genetics to this day \citep[e.g.][]{Coop2009,Akey2010}.

The falling cost of sequencing and genotyping means that such comparisons can now be made on a genome-wide scale, allowing us to start understanding the genetic basis of local adaptation across a broad range of organisms. However, such studies need to acknowledge the sampling issues inherent in population genetic studies of natural populations. In assessing correlations between allele frequencies and environmental variables or looking for loci with unusually high levels of differentiation, two broad technical issues need to be addressed. First, sample allele frequencies are noisy estimates of the population allele frequency, and this issue is exacerbated when sample sizes differ across populations. Second, when multiple populations are compared they are not statistically independent. These populations have experienced varying amounts of shared genetic drift and migration over time and they will consequently vary in their relationship to each other \citep[][]{Robertson1975,Nicholson2002,Excoffier2009, Bonhomme2010}. Failure to account for differences in sample size and the shared history of populations could lead to a high rate of false-positive and negatives due to the unaccounted sources of variance and non-independence among populations. Therefore, accounting for these potential biases should provide additional precision in the identification of loci responsible for adaptation. To accommodate these sources of noise the Bayesian method Bayenv was developed \citep{Coop2010} that attempts to account for these two factors while testing for a correlation between allele frequencies and an environmental variable. To control for a general relationship between populations, a covariance matrix of allele frequencies is estimated from a set of control markers. This model of covariance is then used as a null model against an alternative model which allows for a linear relationship between the (transformed) allele frequencies at a particular locus and an environmental variable of interest. Inference under these models is performed using Markov chain Monte Carlo (MCMC) to integrate over the posterior of the parameters.

Recently, methods closely related to Bayenv have been developed and applied to detect environmental correlations while accounting for population structure.  The most similar approach is by \citet{Guillot2012} who offered large gains in computational efficiency for a model very similar to Bayenv,  but where the covariance matrix has an explicit isolation by distance parametric form, by making use of approximations to perform inference in an MCMC-free framework \citep{Rue2009,Lindgren2011}. \citet{Frichot2012} presented a  Latent Factor Mixed Model that estimates the effect of population history and environmental correlations simultaneously. The \citet{Frichot2012} method resulted in a slightly higher power than Bayenv to detect environmental correlations in simulations, perhaps in part as a result of the simultaneous inference of fixed and random effects reducing the effect of selected loci inflating the covariance matrix. Finally, \citet{Fumagalli2011} and \citet{Hancock2011} used a non-parametric partial mantel test, which makes fewer model assumptions and so should be less sensitive to non-normality. However, the partial mantel test is not well calibrated when both genotypes and environmental variables are spatially autocorrelated \citep[see][for discussion]{Guillot2011}, and so the p-values should be interpreted with caution.

Bayenv has been successfully applied to identify loci putatively involved in local adaptation to environmental variables across a range of different species \citep[e.g.][]{Hancock2008,Hancock2010,Hancock2011a,Hancock2011b,Eckert2010,Fumagalli2011,Jones2011,Cheng2011,Fang2012,Keller2012,Limborg2012,Pyhajarvi2012}. However, further work is needed to make Bayenv robust to outliers and and to extend it to next-generation data applications. One concern about applications of such methods is that linear models are not robust to outliers, which can lead to spurious correlations. For example, if a single population has both an extreme allele frequency and an extreme environmental variable, while all other populations show no correlation, then the linear model may be misled \citep[see ][for examples]{Hancock2011b,Pyhajarvi2012}. This sensitivity can be overcome by using rank-based non-parametric statistics, such as Spearman's $\rho$, which may also offer increased power to detect non-linear relationships. The difficulty is that such tests do not acknowledge the differences in sample size or the covariance in allele frequencies across populations. To overcome these difficulties we provide the user with a set of `standardized' allele frequencies at each SNP, where the effect of unequal sampling variance and covariance among populations has been approximately removed. This affords users a general framework to utilize statistics of their choosing to investigate environmental correlations or other sources of allele frequency variation. As an example of how these `standardized' allele frequencies' can be used we construct a global $F_{ST}$-like statistic that accounts for shared population history and sampling noise. 

We also extend Bayenv to deal with some of the statistical challenges posed by next-generation sequencing. Recently, pooled next-generation sequencing (NGS) of multiple individuals from a population has gained in popularity \citep[e.g.][]{Turner2010,Turner2011,He2011,Kolaczkowski2011,Boitard2012,Fabian2012,Kofler2012,Orozco-Terwengel2012}, as it offers a cost efficient alternative to sequencing of single individuals. However, estimating allele frequencies from read counts sequenced from a pool implies a second level of sampling variance \citep{Futschik2010,Zhu2012}, which needs to be considered in population genetic analyses such as Bayenv. We include the sampling of reads in pooled NGS experiments into the model to account for the additional sampling noise incurred. 

These extensions to Bayenv are implemented in Bayenv2.0 available from \texttt{http://gcbias.org}. We demonstrate the utility of these approaches through simulation and re-analyzing SNP genotyping data from the CEPH Human Genome Diversity Panel \citep[HGDP,][]{Conrad2006,Li2008}.

\section{Methods}
\subsection{General model of Bayenv}

First, we briefly explain the underlying model of Bayenv for the sake of completeness. Further details about the model and inference method can be found in \citet{Coop2010}. Consider a biallelic locus $l$ with a population allele frequency $p_{jl}$ in population $j$ where $n_{j}$ alleles have been sampled from this population in total. We assume that the observed count of allele $1$, $k_{jl}$, in this population is the result of binomial sampling from this population frequency:

\begin{equation}
  P(k_{jl}|p_{j},n_{j})=\binom{n_j}{k_{jl}} {p_{jl}}^{k_{jl}} (1-p_{jl})^{n_j - k_{jl}}. \label{binom}
\end{equation}
 We follow the model of \citet{Nicholson2002} by assuming that a simple transform of the population allele frequency $p_{jl}$ in subpopulation $j$ at locus $l$ represents a normally distributed deviate around an ``ancestral'' frequency $\epsilon_l$. Specifically we assume that 
\begin{equation}
p_{jl} = g(\theta_{jl}) = \left\{ \begin{array}{ll} 0 & \textrm{if } \theta_{jl} < 0 \\
\theta_{jl} &  0\leq \theta_{jl} \leq 1\\
1 &  \theta_{jl} > 1.\end{array} \right.
\end{equation}
i.e. that the mass $<1$ and $>1$ are placed as point masses at $0$ and $1$, representing the loss or fixation of the allele in population $j$ respectively. We then assume that that the marginal distribution of $\theta_{jl}$ is normally distributed, around an `ancestral' mean frequency $\epsilon_l$ with variance proportional to $\epsilon_l(1-\epsilon_l)$ \citep[inspired by the model of][]{Nicholson2002}. We denote the vector of transformed population allele frequencies at a locus by $\theta_l$ where $\theta_l = (\theta_{1l},~\ldots,~\theta_{Jl})$ when $J$ is the number of populations. 
As we do not expect that the populations are independent from each other, we assume that $\theta_l$ follows a multivariate normal distribution
\begin{equation}
  P(\theta_l|\Omega,\epsilon_l) \sim MVN(\epsilon_l,\epsilon_l(1-\epsilon_l)\Omega). \label{MVN_drift}
\end{equation}

We can write the joint probability of our counts at a locus and the $\theta_l$ as
\begin{equation}
 P((k_{1l},\ldots,k_{Jl}),\theta_l |\Omega,\epsilon_l,(n_{1l},\ldots,n_{Jl})) \sim MVN(\epsilon_l,\epsilon_l(1-\epsilon_l)\Omega) \prod_{j=1}^J P(k_{jl}|p_{jl}=g(\theta_{jl}),n_{jl}). \label{jointprob}
\end{equation}

We place priors on $\Omega$ (inverse Wishart) and the $\epsilon_l$ at each SNP (symmetric Beta). Assuming that our SNPs are independent, we write the joint probability of all of our loci and parameters as
\begin{equation}
P(\Omega) \prod_{l=1}^L P((k_{1l},\ldots,k_{Jl}), \theta_l |\Omega,\epsilon_l,(n_{1l},\ldots,n_{Jl})) P(\epsilon_l). \label{jointprob_all}
\end{equation}
Our posterior is this joint probability normalized by the integral over $\Omega$ and the $\epsilon_l$ and $\theta_l$ at all of the loci. 

We then use MCMC to sample posterior draws of the covariance matrix ($\Omega$) from a set of unlinked, putatively neutral control SNPs. Our observations showed that the MCMC converges quickly to a small set of covariance matrices for each data set given a sufficient number of independent SNPs \citep{Coop2010}. Given this tight distribution, we use a single draw of $\Omega$, denoted by $\widehat{\Omega}$, after a sufficient burn in. The entries of the matrix $\Omega$ are closely related to the matrix of pairwise $F_{ST}$ \citep{Weir2002,Samanta2009}, and so this model provides a flexible model of population history; for example \citet{Pickrell2012} used a similar model to infer a tree-like graph of population history and \citet{Guillot2012} uses a related model as a model of isolation by distance.

Next, we formulate an alternative model where an environmental variable $Y$, standardized to have mean $0$ and variance $1$, has a linear effect $\beta$ on the allele frequencies:
\begin{equation}
  P(\theta_l|  \widehat{\Omega},\epsilon_l,\beta, Y) \sim MVN(\epsilon_l+\beta Y,\epsilon_l(1-\epsilon_l) \widehat{\Omega}). \label{alternativemodel}
\end{equation}
To express the support for the alternative model at a locus $l$, \citet{Coop2010} calculated a Bayes factor (BF) by taking the ratio of probability of the alternative and the null model given the data and $\widehat{\Omega}$, integrating out the uncertainty in $\theta_l$, $\epsilon_l$, and $\beta$ (under a uniform prior on $\beta$ between $-0.2$ and $0.2$).

\subsection{Tests based on standardized allele frequencies}
The linear relationship between the transformed allele frequencies (eqn. \eqref{alternativemodel}) may not be the best fit in all situations, as other monotonic relationships (e.g. exponential, logarithmic, saturating) could be viewed as biologically realistic in some cases. Additionally, there may be situations in which a linear model is not robust to outliers and so will spuriously identify loci as strong correlations. Therefore, we provide a general framework to allow investigators to apply statistics of their choice, such as rank-based non-parametric statistics, to detect environmental correlations, while taking advantage of the Bayenv framework. These statistics could in theory be applied to the raw sample frequencies; in practice, however, that can lead to high false positive and false negative rates as sample allele frequencies are naturally noisy because of the process of sampling and non-independent due to the covariance among populations. The multivariate normal framework employed by Bayenv offers a natural way to attempt to standardize $\theta_l$ to be variates with mean zero, variance one, and no covariance. These allele frequencies allow standard statistics that rely on these assumptions to be applied more directly. We denote the Cholesky decomposition of the covariance matrix $C$ ($\Omega=CC^T$, where $C$ is an upper-triangular matrix), which can be thought of as being equivalent to the square root of the matrix, and so analogous to the standard deviation of $\theta_l$.  To standardize the $\theta_l$ for effects of unequal sampling variance, and covariance among populations we write
\begin{equation}
X_l=C^{-1}\frac{(\theta_l - \epsilon_l)}{\sqrt{\epsilon_l(1-\epsilon_l)}}. \label{MVN_standardization}
\end{equation}
If $\theta_l \sim \textrm{MVN}(\epsilon_l, \epsilon_l(1-\epsilon_l) \Omega)$ then $X_l \sim \textrm{MVN}(0,\mathbb{I})$ where $\mathbb{I}$ is the identity matrix (i.e. $\mathbb{I}_{i,j}=1$ if $i=j$ and $\mathbb{I}_{i,j}=0$ otherwise). Note that this transform is not unique, but 
\begin{equation}
X_l^TX_l = \frac{\theta_l^T \Omega^{-1}\theta_l}{\epsilon_l(1-\epsilon_l)} \label{XtX}
\end{equation}
 is. Furthermore, if $\theta_l$ is truly multivariate normal then $X_l^TX_l$ is distributed $\sim \chi^2_J$. This suggests that $X_l^TX_l$ is a natural test statistic to identify loci that deviate away strongly from the multivariate normal distribution, e.g. due to selection. Furthermore, this form naturally accounts for hierarchical population structure, or other models of population structure, that can confound $F_{ST}$-style outlier analyses \citep{Excoffier2009}. Our $X_l^TX_l$ statistic extends the ideas of \citet{Bonhomme2010}, who developed a similar test statistic for the case of a known population tree  (see also \citet{Robertson1975}, for earlier discussion of the effect of a population tree on the \citet{Lewontin1973} test).  

If we wish to test the correlation of our transformed allele frequencies with an environmental variable, we will also need to similarly transform our environmental variable, to ensure that our frequencies and environmental variable are in  the same frame of reference. Specifically if our environmental variable is $Y$ (standardized to be mean zero, variance $1$) then our transformed environmental variable is
\begin{equation}
Y^{\prime} = C^{-1}Y. \label{environ_std}
\end{equation}
Note that this transform will exaggerate the environmental variable difference between very closely related populations. Furthermore, if part of the variation in the environmental variable precisely matches the major of axis of variation in the genetic data, then applying the transform may remove much of this variation. Both of these effects seem desirable properties, as we are interested in identifying correlations discordant with the patterns expected from drift. However, users should visually inspect $Y$ and $Y^{\prime}$ to understand how the transform has altered the environmental variable (see Supplementary Figures 1-4 for examples). 

We do not get to observe $\theta_l$ so we obtain a representative sample of $M$ draws from the posterior ($X_{l}^{(1)},\cdots,~X_l^{(M)}$). Given these draws there is an enormous variety of ways that we could choose to summarize the support for the correlation with our environmental variable $Y^{\prime}$. Here we choose to write 
\begin{equation}
\rho_l(X_{l}^{(1)},\cdots,~X_l^{(M)}) = \frac{1}{M} \sum_{i=1}^M \rho(X_l^{(i)}, Y^{\prime} ), \label{nonparam}
\end{equation}
i.e. $\rho_l$ is the mean of the function $\rho()$ over our posterior draws of $X_l$.

 In the present paper, we calculated Pearson's and Spearman's correlation coefficients (as our $\rho()$) as alternative tests to the Bayes factors. To obtain an appropriate sample from the posterior in a computational efficient manner, these statistics were calculated between $X_l$ and $Y^{\prime}$ every 500 MCMC generations and then averaged over the complete MCMC run. Our draws of $X_l$ will therefore be weakly autocorrelated, but as $\rho_l$ is a mean this does not affect its expectation.

While this standardization, for a known $\widehat{\Omega}$, would work perfectly if our $\theta_l$ were really multivariate normal, in reality this is only an approximation, as even under the null model deviations due to drift are only approximately normal over short time-scales. Thus, while we model drift at a locus as being multivariate normal (i.e. $\theta_l$ has a prior given by eqn. \eqref{MVN_drift}), if the true model is more complex the joint probability of this along with our count data (and our uncertainty in $\Omega$) may force $\theta_l$ to not be $\textrm{MVN}(0,\mathbb{I})$. While, under these circumstances, $X_l$ will conform to those assumptions better than $\theta_l$, we still choose to use the empirical distribution of $\rho_l$ across SNPs rather than rely on asymptotic results, which may not hold. 

\subsection{Sequencing of pooled samples}
If genotyping is conducted as sequencing of population pools, an additional step of sampling is included. At a site $l$ the total coverage of in population $j$ is $m_{jl}$ and we observe $r_{jl}$ reads supporting allele 1. Assuming that each individual contributed the same number of chromosomes to the pool, we can conclude that the sequenced reads are the result of binomial sampling
\begin{equation}
 P\left(r_{jl}|\frac{i}{n_j},m_{jl}\right)=\binom{m_{jl}}{r_{jl}} \frac{i}{n_j}^{r_{jl}} \left(1-\frac{i}{n_j}\right)^{m_{jl}-r_{jl}},
\end{equation}
where $\frac{i}{n_j}$ is the unknown sample allele frequency in the pooled sample. Summing over this unknown frequency 
\begin{equation}
 P\left(r_{jl} |m_{jl},p_{jl},n_{j} \right)=\binom{m_{jl}}{r_{jl}} \sum_{i} \left(\frac{i}{n_{j}}\right)^{r_j} \left(1-\frac{i}{n_{j}}\right)^{m_{jl}-r_{jl}} \binom{n_{j}}{i} {p_{jl}}^{i} \left(1-p_{jl}\right)^{n_{j}-i}
\end{equation}
gives us the probability of our sampled reads given the population frequency. This replaces the binomial probability (eqn. \eqref{binom}) in the joint probability given by eq. \eqref{jointprob}. In \citet{Coop2010} the Bayes factors were approximated by an importance sampling technique while performing MCMC under the null model, i.e. $\beta=0$. This allowed the rapid calculation of the Bayes factor for many environmental variables with little extra computational cost. However, Bayes factors calculated by this technique are noisy, and so here we also implement an MCMC to estimate the posterior on $\beta$. We place a uniform prior on $\beta$ and update $\beta$ along with $\epsilon_l$ and $\theta_l$. For our update on $\beta$ we use a small normal deviate ($\sigma=0.01$) and accept this move with the ratio of the joint posterior of our current parameters to that of our old parameters. As a simple summary of the posterior support for $\beta \neq 0$, we look at the skew of the posterior away from zero. Specifically we estimate the proportion ($f$) of the marginal posterior on $\beta$ that is above $0$, and then take 
$Z=|0.5-f|$ as a test statistic, with values of $Z$ close to 0.5 showing strong support for $\beta \neq 0$.

\subsection{Power simulations}

The extended model was implemented in Bayenv2.0. Simulations were conducted to evaluate the power of these extensions. To use both a realistic covariance among populations and realistic environmental values, we based these simulations on SNP data from the HGDP populations \citep{Conrad2006} and the environmental variables measured at these sampling locations \citep[also used in][]{Hancock2008,Coop2010}. We employed a single Bayenv2.0 estimate of the covariance matrix $\widehat{\Omega}$ from the original SNP data (sampled after 100,000 MCMC iterations) to simulate population allele frequencies. For each SNP, an ancestral frequency $\epsilon_l$ was drawn from a beta distribution (with parameters $\alpha=0.5, \beta=3$). Then population allele frequencies were drawn from the multivariate normal $MVN(\epsilon_i,\epsilon_i(1-\epsilon_i)\widehat{\Omega})$ using the MASS package for GNU R \citep{RDevelopmentCoreTeam2009}. In contrast to the more empirical approach in \citet{Coop2010}, who used observed SNP frequencies, these simulated population frequencies allow us to vary sample size and sequencing coverage for a population. For the simulation of pooled NGS data, we assume that the depth of coverage of a pool follows a negative binomial distribution, which allows for the over-dispersion of read depths compared to the Poisson. Coverages for each population and SNP were independently drawn from a negative binomial distribution $NB(r,p)$ where we set $r=5$ and set $p$ to obtain the respective coverage mean ($NB(r,p)$ has a mean of $pr/(1-p)$ and a variance of $pr/(1-p)^2$). This represents an extreme case where the variance-mean-ratio increases for higher average coverages. Such pattern is generally consistent with observations from pooled next-generation data generated along an altitudinal gradient in \textit{Arabidopsis thaliana} (T.G., C. Lampei, O. Simon and K.J. Schmid, unpublished results). 

To construct a null distribution we calculated Bayes factors or our test statistic $Z$ for these simulated SNPs and an environmental variable $Y$ during 100,000 MCMC iterations. For a second set of SNPs, an environmental effect was simulated by drawing their population allele frequencies from a multivariate normal $MVN(\epsilon_i+\beta Y,\epsilon_i(1-\epsilon_i)\widehat{\Omega})$. Again Bayes factors or our test statistic $Z$ were calculated over 100,000 MCMC iterations. An environmental effect of $|\beta|=0.06$ was simulated when all 52 HGDP populations were used and $|\beta|=0.15$ was used for simulations of smaller population subsets; positive and negative $\beta$s were simulated in identical proportions for all simulations where $Z$ was calculated. To estimate power for our $\rho_l$ statistics samples of 20 chromosomes from each of the 52 HGDP populations were simulated and $\beta$ was varied between 0.01 and 0.09.
Power estimates were based on the proportion of these SNPs that were detected at a certain significance level $\alpha$ ($5\%$ here), i.e. the fraction of our simulations (with a $\beta$) in upper $\alpha$ tail of the null distribution.

\subsection{Data application}

Bayenv2.0 was used to re-analyze a genome-wide data set of 640,698 SNPs from 52 HGDP-CEPH populations \citep{Li2008,Hancock2010} using Bayes factors and our non-parametric test statistic ($\rho_l$). We restricted our analysis to winter conditions, as most winter climate variables show outliers and a non-normal distribution. All environmental variables were normalized to a mean of zero and a standard deviation of one. The covariance matrix was estimated from a random subset of 5,000 SNPs after 100,000 MCMC iterations. Bayes factors and correlation coefficients for each SNP were estimated during 100,000 MCMC iterations. In addition to these test statistics, we sampled $X_l$ every 500 MCMC generations and calculated $X_l^T X_l$. These values were averaged per SNP to calculate $\overline{X_l^TX_l}$ and to check for deviations from the multivariate normal distribution for each SNP individually. SNP positions and gene annotations were obtained from Ensembl \citep{Flicek2012} and Entrez Gene \citep{Maglott2011}. 

\section{Results}

\subsection{Using tests based on standardized allele frequencies}
\begin{figure}[h]
 \centering
 \includegraphics[width=\textwidth]{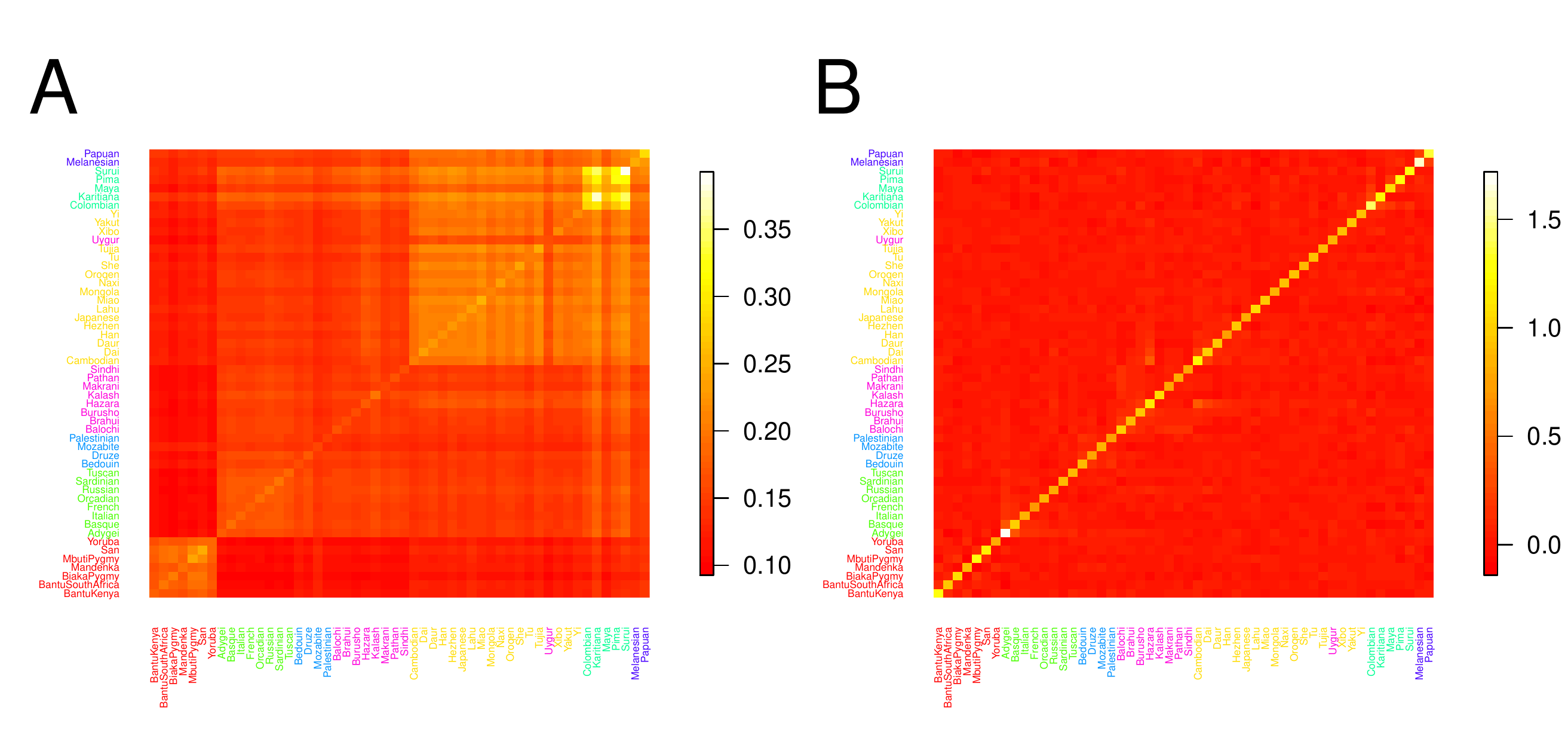}
\caption{Covariance among HGDP populations estimated by Bayenv2.0 and the covariance calculated on the $X$s for the same SNPs. Populations are colored according to broad geographic regions used in \citet{Rosenberg2002}.}\label{img:covariance}
\end{figure}

We explored the performance of tests based on our standardized transformed population frequencies ($X_l$). Before we calculate test statistics on our standardized allele frequencies, we first examined whether the multivariate standardization (as in eqn. \eqref{MVN_standardization}) had removed the covariance among populations from our standardized $X_l$. We first calculated the sample covariance matrix using the sample frequencies for 2,333 HGDP SNPs \citep[the dataset of][]{Conrad2006} shown in Figure~\ref{img:covariance}A. Specifically, denoting the vector of sample frequencies by $k_{l}/n_l$ we calculated  $\frac{1}{L}\sum_{l=1}^{L} (k_{l}/n_l)(k_{l}/n_l)^T$. As expected, there is substantial structure in this sample covariance matrix between regions, which corresponds to known population structure \citep{Coop2010}. Then we calculated the sample covariance matrix of the $X_l$ across these SNPs using Bayenv2.0; specifically we took a single draw of $X_l$ (after a burnin) for each of these 2,333 SNPs, and calculated $\frac{1}{L}\sum_{l=1}^{L} X_l X_l^T$. The resulting sample covariance matrix (shown in Figure~\ref{img:covariance}B) is close to the identity matrix in form, demonstrating that the majority of the covariance between populations has been removed. This suggests that our $X_l$ are appropriately standardized for the application of correlation tests averaging across our uncertainty in $X_l$ at each locus.

\begin{figure}[hp]
 \centering
 \includegraphics[width=\textwidth]{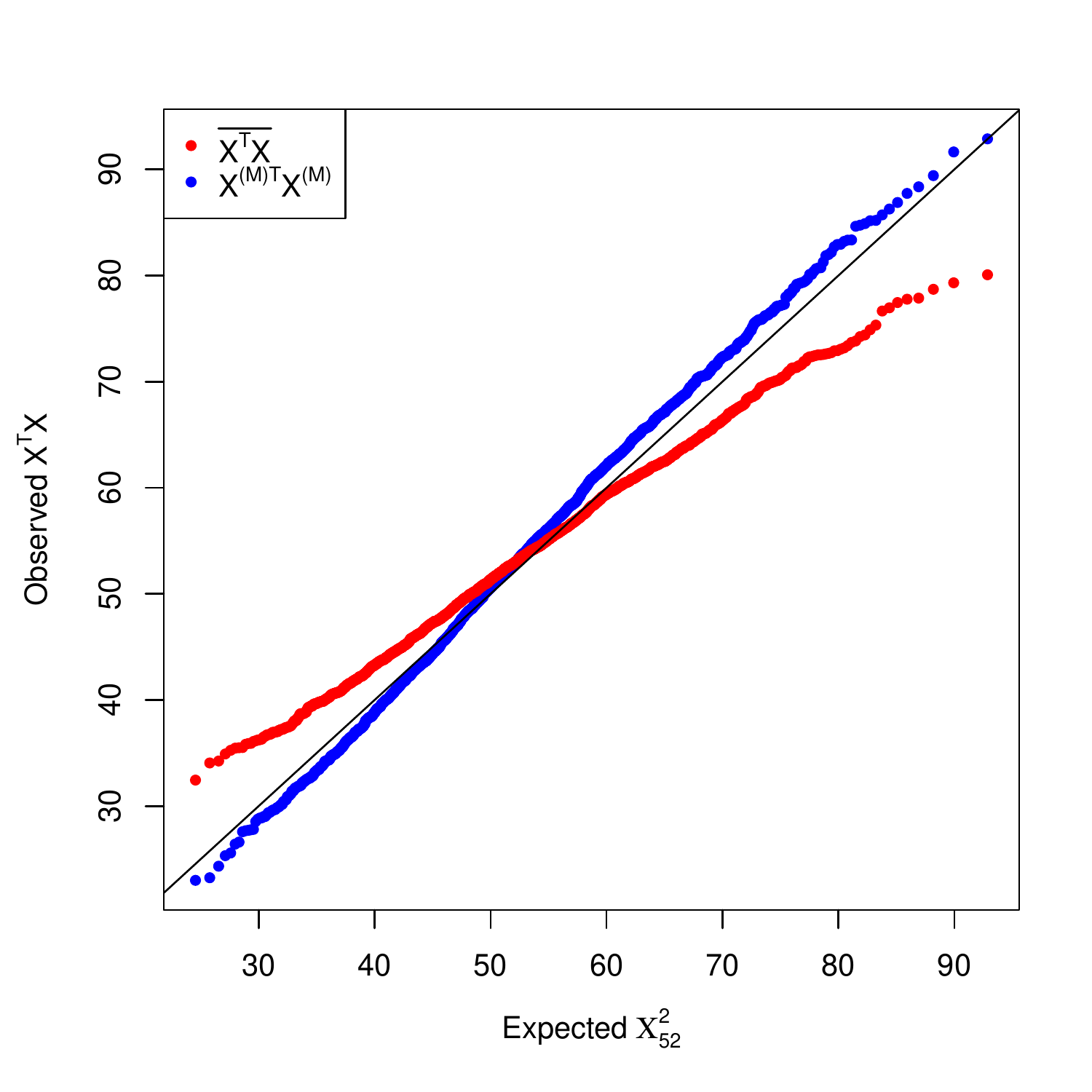}
\caption{QQ-plot of $X_l^TX_l$ calculated in two different ways and the $\chi_{52}^2$ distribution, which is expected if $X_l$ would follow $\textrm{MVN}(0,\mathbb{I})$.}\label{img:qq}
\end{figure}

 To further test the normality of $X_l$, we checked if $X_l^TX_l$ follows a $\chi^2$ distribution with 52 degrees of freedom (i.e. the number of populations, see Methods). To test this, $X^TX$ was calculated in two different ways, first using the final generation of the MCMC ($X_l^{(M)T}X_l^{(M)}$) and the second using the average $X_l^TX_l$ across all $M$ samples for each locus $l$ ($\overline{X_l^TX_l}$). Figure \ref{img:qq} shows a QQ-plot of the $X_l^TX_l$ and the expected $\chi_{52}^2$ distribution.
The mean of each distribution approximately matches that of the $\chi^2_{52}$, whereas the variances do not. The estimates based on single samples from the MCMC show a somewhat higher variance. The averaging, on the other hand, led to a smaller variance, indicating that this approach is slightly over-conservative. Both observed distributions are not consistent with the expected $\chi_{52}^2$ distribution (Kolmogorov-Smirnov tests, both p-values $<10^{-6}$). Therefore, while $X_l^T X_l $ provides a potentially suitable summary statistic for identifying empirical outliers, we cannot assume a distributional form to those outliers under a null neutral model. We chose to use $\overline{X_l^TX_l}$, as it averages over our uncertainty in the sample frequencies, and so should be more robust to outliers due to small sample sizes.

\begin{figure}[htb]
 \centering
 \includegraphics[width=\textwidth]{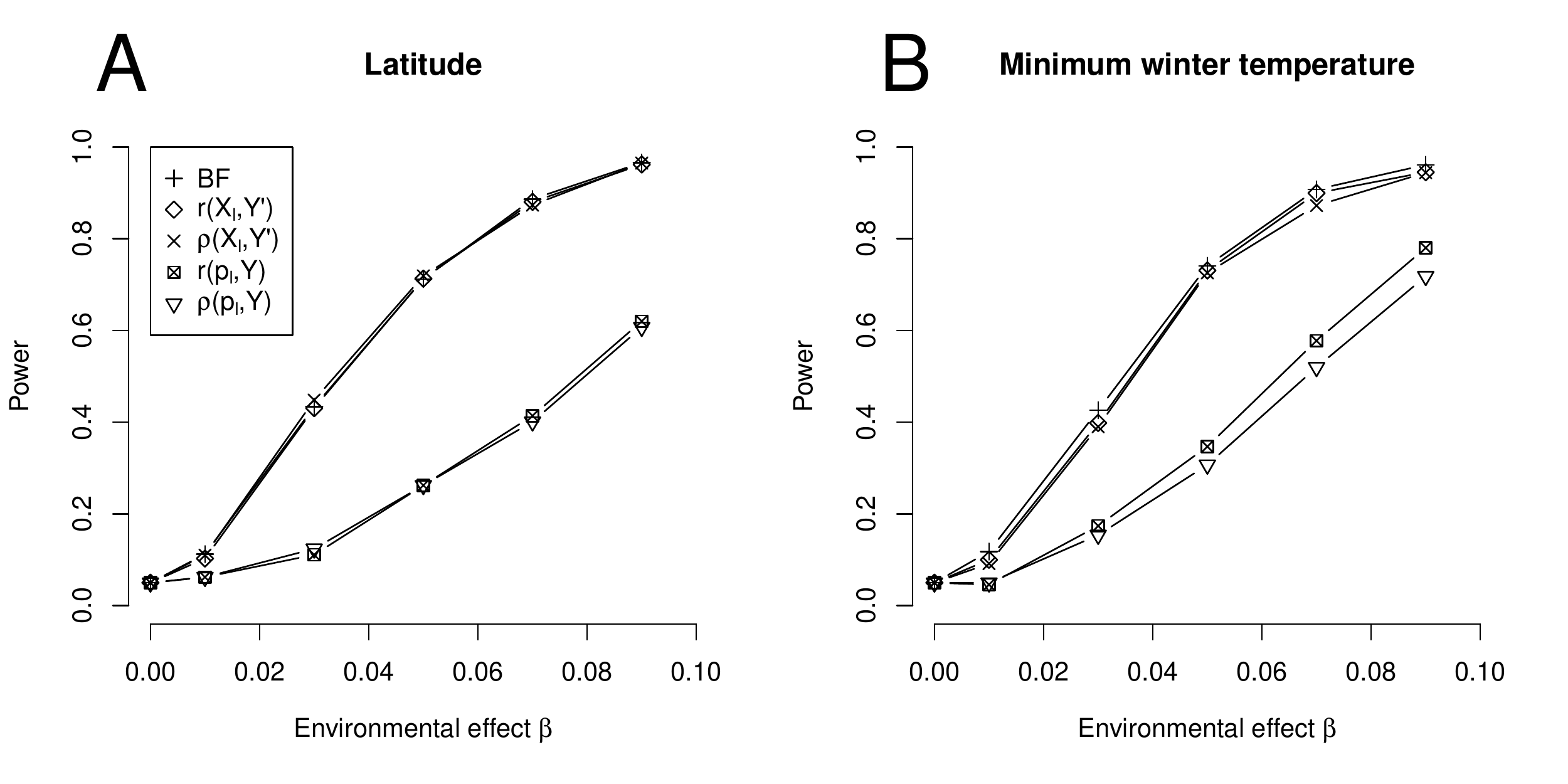}
\caption{Power of Bayes factors compared to power of correlation coefficients on X based on simulations for all 52 HGDP populations for the environmental variables latitude (A) and minimum winter temperature (B).}\label{img:correlation_power}
\end{figure}

To explore the power of standard correlation tests applied to our standardized $X_l$, in comparison to the Bayes factors, we again conducted power simulations based on the HGDP data. We also calculated both Spearman's $\rho$ and and Pearson's $r$ 
between $Y^{\prime}$  and our transformed allele frequencies averaged across the posterior on these transformed frequencies. We transformed our latitude and minimum winter temperature value, our $Y$'s, to give us $Y^{\prime}$ as in eqn. \eqref{environ_std} (see Supplementary Figures 1-4). 
Statistics based on our Bayesian model clearly outperform correlation tests calculated for point estimates from sample allele frequencies (Figure~\ref{img:correlation_power}). This improvement in power is due to the fact that the methods based on the sample frequencies fail to incorporate the sampling noise and the relationship among populations. All three tests based on Bayenv performed effectively identically with marginal advantages of the Bayes factors for minimum winter temperature (Figure~\ref{img:correlation_power}B) and a slightly lower power of Spearman's $\rho$, which is not surprising, as all simulated effects are linear. We expect that the relative performance of the rank-based test, i.e. Spearman's $\rho$, may be reduced as the number of populations is decreased. We also tested the power of the  $X_l$ tests incorrectly using $Y$ in place of $Y^{\prime}$; this gave rise to power curves intermediate between the two sets (data not shown). Overall these results show that correlation tests based on $X_{l}$ perform well. 

The alternative model of Bayenv (eqn. \eqref{alternativemodel}) implies a linear relationship between the transformed allele frequencies and the environmental variable. However, the fitting and significance of this linear model may be misled by populations that are statistical outliers. For instance, linear models might mistakenly identify cases as strong candidates, when allele frequencies and environment for all but one population are consistent with our null model and this single outlier population features both an extreme environment and an extreme allele frequency. We note that the extreme allele frequency may be due to a component of drift not well modeled by our MVN framework, or due to a selection pressure (or response) poorly correlated with our environmental variable of interest. While loci of the latter form are of interest as \emph{genomic} outliers, we believe researchers interested in particular environmental variables would consider such loci spurious, and would prefer a set of candidates where many populations support a consistent pattern.

\begin{figure}[htb]
 \centering
 \includegraphics[width=\textwidth]{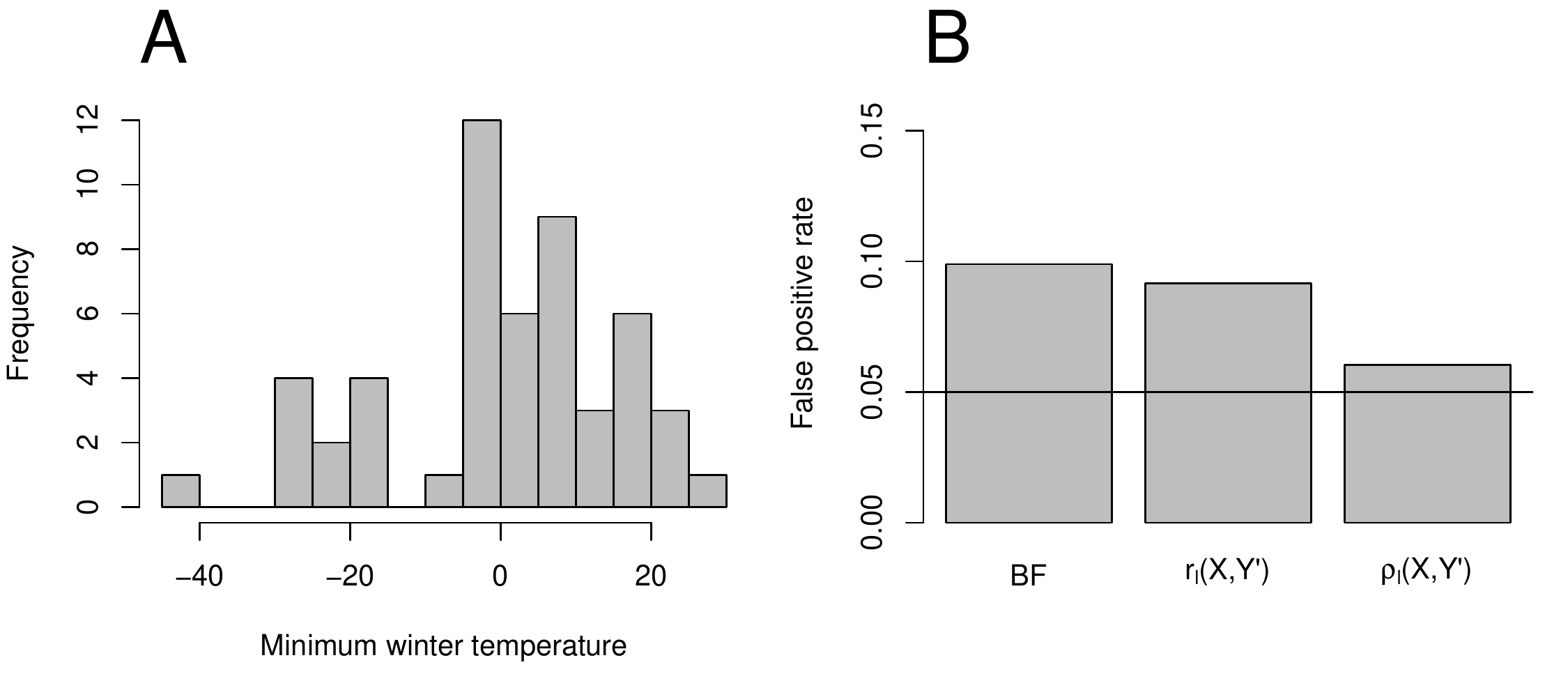}
\caption{False-positives induced by populations at extreme conditions. (A) Histogram of minimum winter temperature for the 52 HGDP populations. (B) False-positive rate of different statistics if one allele is fixed in the coldest population.}\label{img:fool}
\end{figure}

To test such a case, we simulated allele frequencies for the HGDP populations based on $MVN(\epsilon_l,\epsilon_l(1-\epsilon_l)\widehat{\Omega})$ as described above. Winter minimum temperature was used as climate variable since one population, the Yakuts from north-east Russia, is characterized by a very low minimum temperature (Figure~\ref{img:fool}A). To create outliers, we set the allele frequencies of the Yakuts to 0. Both statistics based on linear models, Bayes factors and Pearson's correlation coefficient $r$ showed an excess of false positives (Figure~\ref{img:fool}B), while a non-parametric statistic, in this case Spearman's rank correlation coefficient $\rho$, was much less sensitive to these outliers, with a false-positive rate very close to the expected value of $5\%$ (Figure~\ref{img:fool}B). 

\subsection{Simulation of pooled data}

Pooled sequencing of multiple individuals has increased in popularity, as it is considerably cheaper than barcoding all individuals and sequencing them separately \citep[but see][]{Cutler2010}. The use of allele frequencies estimated from the resulting read counts seems to be a reasonable application of our method. However, it raises the question how Bayenv behaves for different coverages as increasing sequencing coverage is not the same as increased numbers of sampled individuals. Therefore, we simulated data that resembles the HGDP populations and then pooled 10 diploid individuals (i.e. 20 chromosomes) from each population and used the populations' respective latitudes as our environmental variable.

\begin{figure}[hp]
 \centering
 \includegraphics[width=\textwidth]{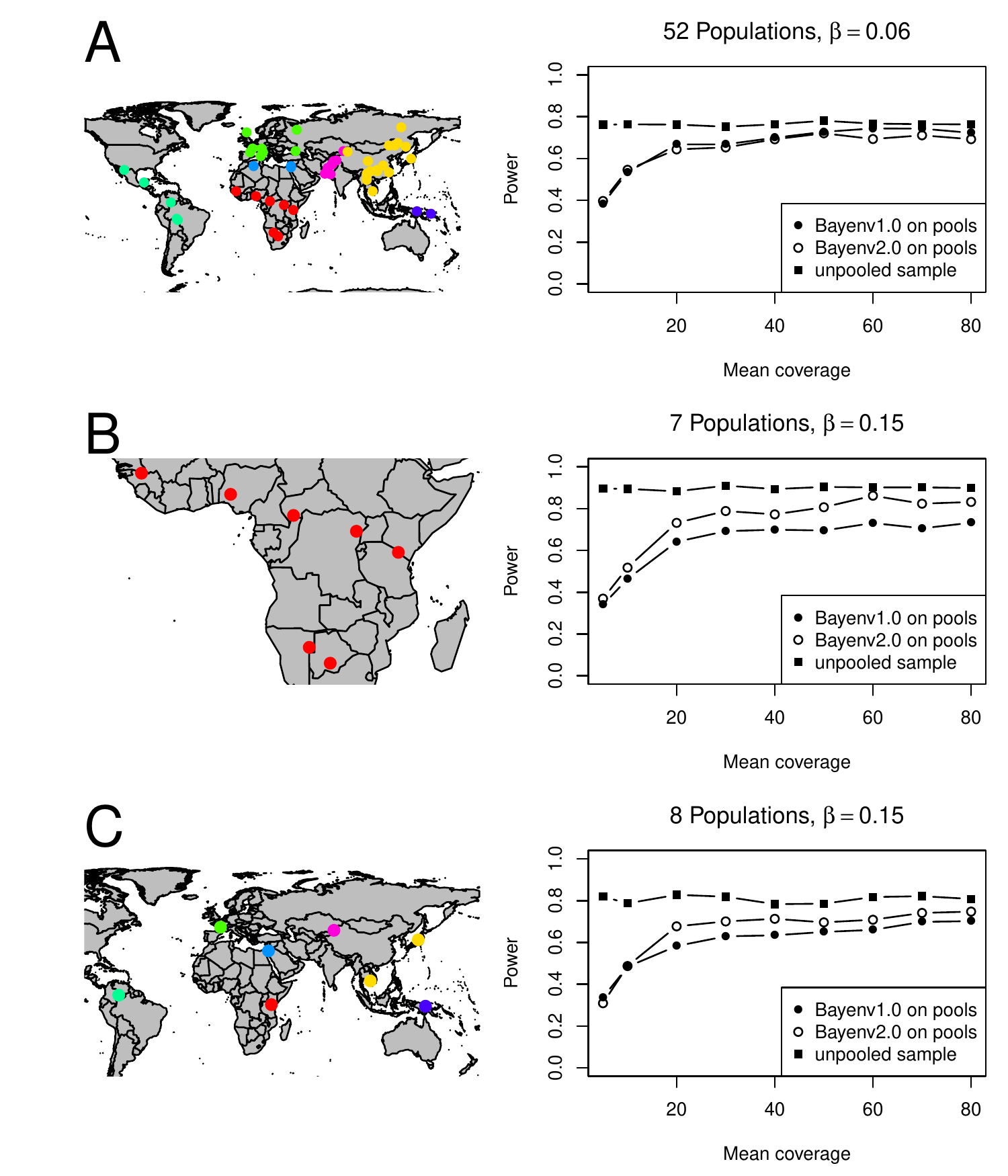}
\caption{Power to detect environmental correlations with latitude in pooled samples. (A) in all 52 HGDP populations, (B) in the seven sub-Saharan populations and (C) in one population per broader geographic region. Populations are colored as in Figure~\ref{img:covariance}.}\label{img:nbinom_power}
\end{figure}

We first experimented with incorrectly using read counts in place of the chromosome counts (i.e. assuming $r_{jl}$ and $m_{jl}$ were $k_{jl}$ and $n_{jl}$, respectively), and found that this resulted in an excess of extreme Bayes factors for high coverages under the null (data not shown). We found this inflation to be most pronounced when read depths are greater than the actual sample size, and this is likely due to false certainty about the population frequencies. 
We then ran power simulations of Bayenv matched to the HGDP data, using $Z$ as a test statistic, with the true sample frequencies (black squared in Figure~\ref{img:nbinom_power}), and incorrectly using the read counts as the input data for the previous version of Bayenv (Bayenv1.0, black circles in Figure~\ref{img:nbinom_power}). Bayenv2.0, which accounts for both stages of binomial sampling in pooled data (as described above), was also applied to the same read counts (white dots in Figure~\ref{img:nbinom_power}). The true sample frequencies naturally resulted in the best power as there is no additional sampling noise (Figure~\ref{img:nbinom_power}A). For higher mean coverages the power of Bayenv1.0 using the read counts as sample allele frequencies was almost as good as the power using true sample allele frequencies (Figure~\ref{img:nbinom_power}A). As most applications may consist of a smaller number of populations, we additionally sampled two subsets consisting of all HGDP sub-Saharan African populations (7 populations; Yoruba, San, Mbuti Pygmy, Mandenka, Biaka Pygmy, Bantu South Africa, Bantu Kenya; Figure~\ref{img:nbinom_power}B) and eight populations spread over the entire globe (Bantu Kenya, French, Bedouin, Cambodian, Japanese, Uygur, Colombian, Papuan; Figure~\ref{img:nbinom_power}C). On all of these, Bayenv using the true sample frequencies out-performed Bayenv1.0 using the read counts.

In part the poor power in pooled studies is unavoidable due to the additional sampling noise. However, the loss of power is likely boosted by failing to properly account for this second stage of sampling, which leads to poor performance due to variation in depth across populations and SNPs. The extended model of Bayenv2.0 should compensate the loss of power to some extent. Somewhat surprisingly, we did not observe any advantages of the extended model in detection power if all 52 HGDP populations are simulated (Figure~\ref{img:nbinom_power}A). The small differences between the extended model and incorrectly using the read counts as the input are mainly due to convergence of the MCMC, which is somewhat slower incorporating both levels of sampling. Including the sampling of reads into the model had a clearly positive effect on power in our population subsets and incorrectly using the read counts as input did not reach similar powers for high coverages (Figure~\ref{img:nbinom_power}B, C). However, power of Bayenv2.0 was still considerably low for mean coverages $<20\times$, suggesting that such low read depths do not provide enough certainty for reliable frequency estimation. 

Notably, a simulated effect of identical magnitude was detected with a higher power in the seven sub-Saharan populations than in the eight worldwide populations. Additionally, the power difference between the extended model and incorrectly using the read counts as the input was higher in the sub-Saharan populations. This demonstrates the effect of different covariance structures among populations (see Figure~\ref{img:covariance}A) and their relation to the environmental variable on the performance of Bayenv. Presumably this lower power in the world-wide samples is due to the fact that the levels of drift are higher between world-wide populations than within Africa.

\subsection{Robust candidates in the HGDP data}
Finally we explored the use of our standardized $X_l$ for identifying robust putative candidates for adaptive evolution in the HGDP data of \citet{Li2008}.

As described above, populations with outliers in terms of allele frequencies and/or environments can potentially lead to spurious correlations. For example, the use of minimum winter temperature as an environmental variable could generate false positive correlations in analyses of the HGDP data because of the extremely low temperature for the Yakut population. To explore this, we used minimum winter temperature and re-analyzed all 640,698 SNPs of the HGDP data, calculating both Bayes factors and $\rho_l(X_l,Y^{\prime})$ using Spearman's rank correlation coefficient $\rho$. Our Bayes factors and $|\rho_l(X_l,Y^{\prime})|$ are correlated across SNPs (Spearman's $\rho=0.72$) and show an overlap of 29 SNPs in their top 100 most extreme SNPs, 142 SNPs in the top 500 and 2.8 \% in the top 5 \% signals. These overlaps are substantial but suggest that our two tests are detecting somewhat different signals, which likely reflects in part the influence of outlier populations. 

The 100 strongest signals of the Bayes factor analysis and $|\rho_l(X_l,Y^{\prime})|$ are shown in Supplementary Tables 1 and 2. The top 5 Bayes factors include SNPs that fall in potential candidate genes, such as epidermal growth factor receptor  \citep[\textit{EGFR},][]{Hancock2008} and a non-synonymous SNP in zonadhesin \citep[\textit{ZAN},][]{Gasper2006}, both of which were previously identified in small scale selection scans. We also find a SNP (rs6500380) located in a region associated with earwax type (i.e. wet or dry) which has been subject to a selective sweep in East Asian populations \citep{Ohashi2011}. Further signals fall in genes involved in fat metabolism, which is a plausible trait for the adaptation to low temperatures. Among our top hits multiple SNPs fall in the gene \textit{MKL1} (megakaryoblastic leukemia 1), which is a myocardin-related transcription factor that has been associated with various disease phenotypes \citep{Ma2001,Hinohara2009,Scharenberg2010}, but is also involved in smooth muscle cell differentiation, mammary gland function, and cytoskeletal signaling \citep{Parmacek2007,Maglott2011}.

\begin{figure}[hp]
 \centering

 \includegraphics[width=\textwidth]{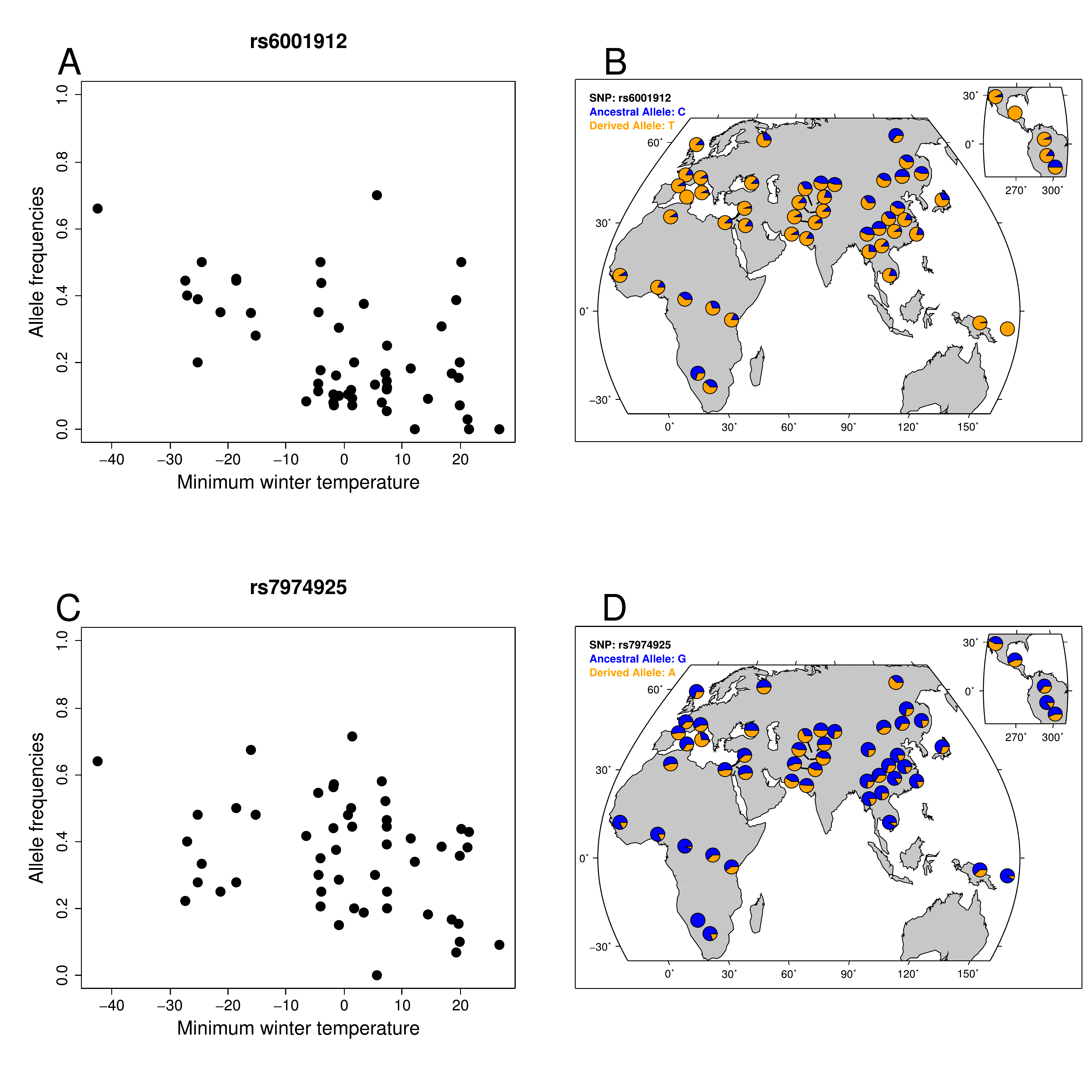}
\caption{Two exemplarily chosen SNP from the top 20 Bayes factors. (A) Allele frequencies and standardized minimum winter temperatures of rs6001912 which is among the top 25 SNPs of both statistics BF and $\rho$, (B) shows the geographical distribution of rs6001912. (C) rs7974925 is among the top 20 BFs but only the top 7,000 $\rho$ signals which is mainly caused by the two outlier populations, (D) shows the geographical distribution of rs7974925. Plots of geographic distributions were downloaded from the HGDP selection browser \citep{Pickrell2009}.}\label{img:snp_example}
\end{figure}

To exemplify the effect of an outlier, we compare two SNPs that fall in our top 20 Bayes factors. Both SNPs, rs6001912 and rs7974925 (Figure~\ref{img:snp_example}A, C), are characterized by similarly high Bayes factors (Supplementary Table~1) and extreme allele frequencies in the Yakuts (Figure~\ref{img:snp_example}B, D). However, only rs6001912 is among the top 25 signals for both statistics, whereas rs7974925 is only among the top 5 \% of Spearman's $\rho$ (Supplementary Tables~1, 2). This suggests that the Bayes factor signal at rs7974925 is strongly driven by the low allele frequency in the Yakuts, and the signal at rs6001912 is more robust even without this outlying data point (Figure~\ref{img:snp_example}A, C). We suggest that the Bayes factors, or other linear model test statistics, should be used in conjunction with robust test statistics such as those described here to avoid spurious signals due to outliers. As these both can be calculated from the same MCMC run, this should be reasonably computationally efficient.

\begin{figure}[hp]
 \centering
 \includegraphics[width=\textwidth]{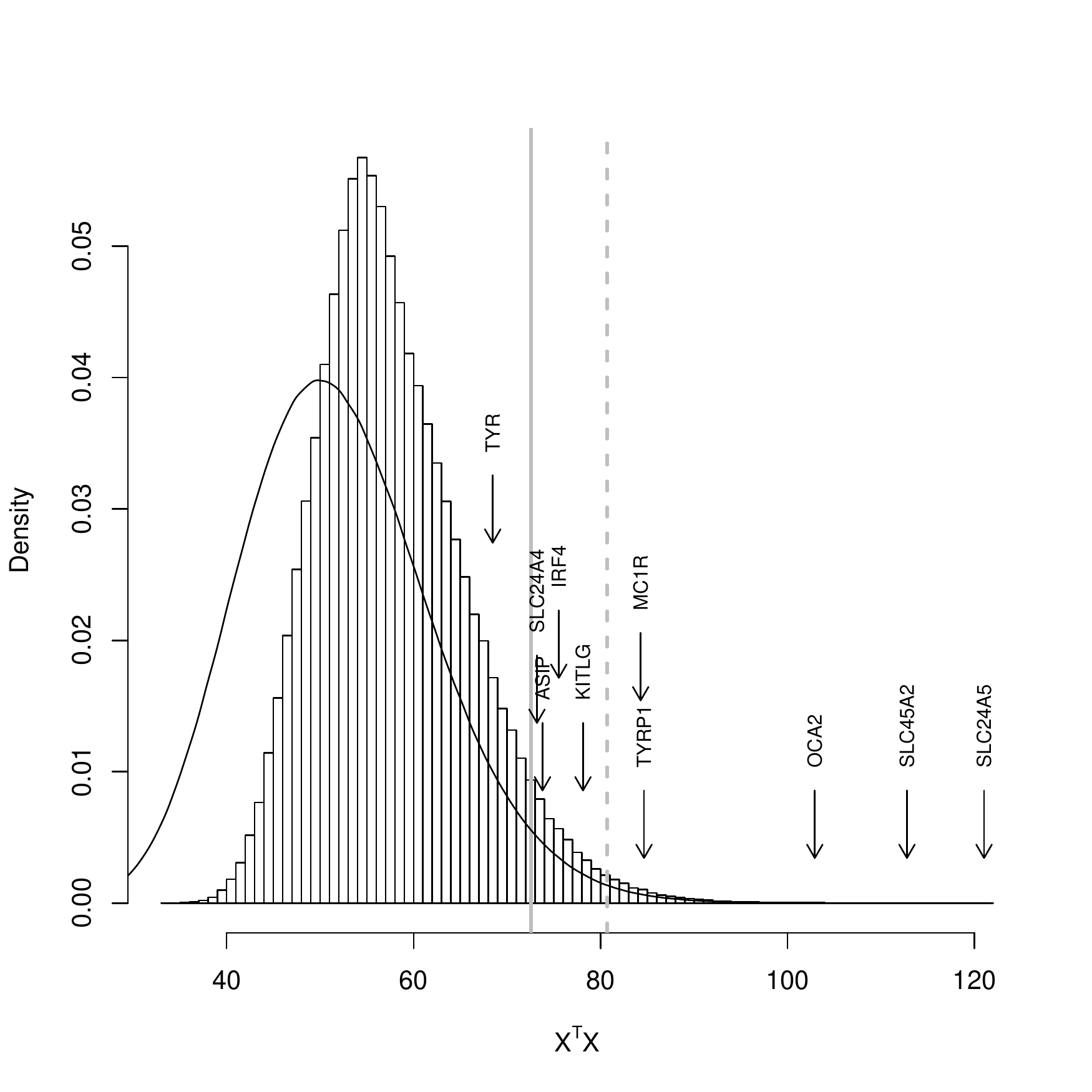}
\caption{Histogram of $X_l^TX_l$ calculated for all HGDP SNPs. The labels of candidate genes are shown at the maximum $X_l^TX_l$ of any SNP within 50kbp up- and downstream of the particular gene. The solid line shows the position beyond which 5\% of all $X_l^TX_l$ fall, the dashed line denotes the top 1\%. The solid black line shows the density of the expected $\chi_{52}^2$ distribution.}\label{img:xtx_hist}
\end{figure}

We also explored our test statistic  $\overline{X_l^TX_l}$, designed to highlight loci that deviate strongly from the expected pattern of population structure, calculated for each of the 640,698 HGDP SNPs. These have been uploaded as a genome browser track to \texttt{http://hgdp.uchicago.edu/}. The empirical distribution is shown in Figure~\ref{img:xtx_hist}. The empirical distribution clearly differs from the expected $\chi^2_{52}$ distribution, having a higher mean and a lower variance than expected. This again highlights that we do not have a good theoretical expectation for the distribution and so must use the empirical ranks to judge how interesting a signal is. To briefly explore where known signals fall in our empirical distribution in Figure~\ref{img:xtx_hist}, we also plot as arrows the maximum  $\overline{X_l^TX_l}$ for SNPs that fall within 50 kbp up- and downstream of ten well known pigmentation genes \citep[list taken from][]{Pickrell2009}. As these arrows represent maximums across a number of SNPs around the gene, they will necessarily be more extreme than an average draw from this distribution. However, the extreme signals at a number of these genes demonstrate that the method is detecting loci with extreme allele frequency patterns. 
The SNP with the most extreme value of $\overline{X_l^TX_l}$ in the genome falls close to \textit{SLC24A5} \citep{Lamason2005}, with a SNP close to \textit{SLC45A2} being the second largest signal in the genome \citep{Nakayama2002}. More generally, five of these ten pigmentation genes fall in the top 1\% and nine genes fall in the top 5\% of the empirical distribution. A SNP close to the gene \textit{EDAR}, one of the highest pairwise $F_{ST}$ between East Asia and Western Eurasia HGDP populations, is also in the top ten SNPs \citep{Sabeti2007}. 

To examine the relationship between $\overline{X_l^TX_l}$ and global $F_{ST}$ we took per SNP values of global $F_{ST}$ previously calculated among the colored groupings depicted in Figure~\ref{img:covariance} \citep[values from][]{Pickrell2009, Coop2009}. The Spearman's $\rho$ between $\overline{X_l^TX_l}$ and  $F_{ST}$ was $0.48$. Looking at the extremes of both distributions, $\overline{X_l^TX_l}$ and $F_{ST}$ show an overlap of 6 SNPs in their top 100 most extreme SNPs, 37 SNPs in the top 500 and 1.4 \% in the top 5 \% signals. In Supplementary table 3 we present the top 100  $\overline{X_l^TX_l}$ SNPs in the genome, along with their nearest genes and global $F_{ST}$ values. 

The weak overlap in the tails of the genome-wide $\overline{X_l^TX_l}$ and $F_{ST}$ means that they are finding different sets of candidate SNPs, presumably due to the reweighting of allele frequencies in $\overline{X_l^TX_l}$. For example, our $12^{th}$ highest SNP for  $\overline{X_l^TX_l}$ falls close to \textit{MCHR1}, with our $21^{st}$ highest gene being a non-synonymous variant (rs133072) in this gene. \textit{MCHR1} (Melanin-concentrating hormone receptor 1) is known to play a role in role in the intake of food, body weight, and energy balance in mice \citep{Marsh2002}, and the effect of the nonsynonymous variant on obesity has been debated \citep{Wermter2005,Rutanen2007,Kring2008} but the variant did not achieve genome-wide significance in a large genome-wide association meta-analysis of BMI \citep{Speliotes2010}. Both of these SNPs are nearly fixed differences between East Asia and the American HGDP populations (Supplementary Figures 5 and 6). This strong difference between regions that share a recent history and, thus, covariance among allele frequencies (Figure~\ref{img:covariance}), makes these SNPs an interesting pattern for $\overline{X_l^TX_l}$. However, neither of the two SNP has an extremely impressive global $F_{ST}$ (falling in only the $5\%$ tail), presumably because East Asia and the American HGDP populations are only two of seven groups in the global $F_{ST}$ calculation and the other five groups do not show an interesting pattern.

\section{Discussion}
In this article we have presented a method to more robustly identify loci where allele frequencies correlate with environmental variables.  We have also described a method to detect loci that are outliers with respect to genome-wide population structure, while accounting for the differential relatedness across populations.

Many available tests for selection are designed to detect rapid complete sweeps from new mutations; however, such events are likely just a small percentage adaptive genetic change \citep{Coop2009,Pritchard2010, Cao2011,Hernandez2011}. Analyzing allele frequencies across multiple populations offers the opportunity to detect selection acting on standing variation and polygenic phenotypes. The falling cost of genotyping means that typing individuals from many populations is now in reach, which will allow us to connect environmental variables to more subtle adaptive genetic variation. However, we stress that loci detected by the approaches discussed above are obviously at best just candidates for being involved in adaptation to a particular climate variable, or set of climate variables, and so additional evidence is needed to build the adaptive case at any locus.

Our use of the covariance matrix of population allele frequencies when looking for environmental correlations is conceptually similar to linear mixed model (LMM) approaches that account for kinship structure in genome-wide association studies (GWAS) \citep[e.g.][who use a observed relatedness matrix as the covariance matrix of the random effect]{Yu2006,Kang2008,Kang2010,Zhou2012}. One important difference is that we seek to predict allele frequencies at a locus using the environmental variable, whereas these LMM methods are predicting a phenotype as a function of genotypes at a locus. In our approach the equivalent of the random effect matrix is a proxy for a neutral model of allele frequency variation, while in the application to GWAS the kinship matrix accounts for the cofounding due to heritable variation in the phenotype elsewhere in the genome. Our model could be used to detect loci that were strongly covaried with population mean phenotypes \citep[e.g. phenotypes measured at the breed level in dogs][]{Boyko2010}. However, the method used this way would have a high rate of false positives if there are large environmental effects on the phenotype that coincide with the principal axes of the covariance matrix. Similarly, LMM approaches could be used to identify loci which covaried with environmental gradients, but they may be underpowered as their random effects model does not attempt to reflect a model of genetic drift.

\paragraph{Standardized allele frequencies}
We introduced a set of tests based on using our model of the covariance of allele frequencies to produce a set of standardized allele frequencies ($X_l$). The calculation of standardized allele frequencies allows us to calculate a variety of statistics while taking advantage of the other features of Bayenv2.0's approach to account for covariance among populations and sampling noise. The removal of covariance is often a standard step in multivariate analysis; here we remove this covariance structure in a way that acknowledges the approximate form of genetic drift and the bounded nature of allele frequencies. By integrating our statistic across the posterior for $X_l$, we are averaging across our uncertainty in allele frequencies, which should further increase our power.   

As an example of the usefulness of the $X_l$, we explored their application in identifying robust correlations with environmental variables. While the use of Spearman's $\rho$ on these transformed allele frequencies results in a small loss of power, it is much less sensitive to outliers and able to detect any monotonic relationship. Therefore, a combined approach which takes a set of SNPs in the intersection of the tail of Bayes factors and in the tail of Spearman's $\rho$ on our transformed allele frequencies should provide best results.  

Our transformed allele frequencies could also be used to detect and distinguish between the effects of multiple environmental variables shaping variation at a locus. This could be accomplished by including the multiple transformed environmental variables ($Y^{\prime}$) into a linear model to predict the $X_l$ at a locus or by applying appropriate transformed ecological niche models (ENM) to the $X_l$ to understand the predictors of allele frequencies at a locus \citep[see][for applications of ENMs to allele frequencies]{Fournier-Level2011,Banta2012}. However, there is limited information about the effects of even a single environmental variable from contemporary allele frequencies if neutral allele frequencies are autocorrelated on the same scale as environmental variation (as is the case in humans). Therefore, we caution that in many situations there will be very limited power to learn about the effect of multiple environmental variables. 
 
Using our $X_l$ statistics, we also introduced a method to identify loci that are outliers from the general pattern of population structure (our $X^TX$ statistic). This statistic is closely related to $F_{ST}$, which can be expressed as $Var(p_{lj})/ \left( \epsilon_l (1- \epsilon_l) \right)$, where $Var(p_{lj})$ is the variance of our allele frequency across populations \citep[see][for discussion]{Nicholson2002, Balding2003, Bonhomme2010}.  Our statistic, which is the variance of $X_l$, can be written as eqn. \eqref{XtX}, and so $X^T X$ can be seen as closely related to calculating $F_{ST}$ on the standardized allele frequencies. Importantly, by removing the covariance, we reweight populations so that a small change shared across many closely related populations is downweighted. This reweighting therefore should increase our power to detect unusual allele frequencies compared to global $F_{ST}$. The fact that we remove the covariance between closely related populations also means that, unlike $F_{ST}$-based methods, we do not have to arbitrarily clump populations in order to identify globally differentiated SNPs. While in this paper we use the 52 HGDP population labels, in principle Bayenv2.0 could be run treating each individual as a population, allowing $X^T X$ to be calculated without regard to any population label. However, this would be computationally time-consuming with thousands of individuals. In that case perhaps the sample frequencies and the sample covariance matrix, could instead be used to mitigate the computational burden.    

Ideally our $X^TX$ statistic would have a parametric distribution under a general null model where only drift and migration shaped our frequencies. That might allow us to make statements about what fraction of allele frequency change was due to selection. Indeed, as noted above, if our population frequencies were truly multivariate normal, our $X^TX$ statistic would be $\chi^2$ distributed if our sample sizes were sufficiently large. This assumption would be approximately met if our levels of drift were sufficiently small, such that the transition density of allele frequencies was well approximated by a normal \citep[see][for recent empirical applications along these lines]{Price2009,Bhatia2011}. However, when levels of drift are higher, our normal approximation will be break down, as demonstrated by the poor fit of the $\chi^2$ to the transformed HGDP frequencies. The distribution of our statistic could be obtained by simulation if the population history were known. In practice, we are skeptical that our knowledge of population genetic history will be sufficiently accurate to make this feasible, but simulations may be useful in guiding the setting of approximate significance levels.   

\paragraph{Pooled Next-generation sequencing}
Recent empirical validations have shown that pooled re-sequencing of populations is a powerful and cost-efficient way to estimate allele frequencies \citep{Zhu2012}, but see \citet{Cutler2010}. The down-side of the saving of costs in library preparation and sequencing is the potential for increased sampling noise in the allele frequency estimates \citep{Futschik2010,Zhu2012} and the loss of haplotype information \citep[although some haplotypic information can be recovered,][]{Long2011}. We account for the sampling of sequencing reads as an additional level of binomial sampling in the model of Bayenv2.0. Our power simulations show that accommodating the extra level of sampling in pooled designs can help to improve the power. However, they also highlight the large unavoidable loss in power due to increased sampling noise when the depth of coverage is low. The only way that this can be circumvented is through increasing sequencing coverage to provide sufficient certainty in the estimated allele frequencies and, thus, sufficient power to detect environmental correlations. Although low fold sequencing of many populations may help to increase power in some situations, is likely that for some species (notably humans) sampling, and not sequencing, will be the limiting resource in the future.

Our model of pooled resequencing in Bayenv2.0 implies uniform sampling of reads from each individual. Therefore, we do not account for the possibility of an unequal number of chromosomes per individual due to measurement errors, different DNA content per individual, or differences caused during DNA extraction, all of which might cause additional noise in the allele frequency estimation \citep{Futschik2010,Cutler2010}. This additional noise, if it is constant across loci, should be absorbed into the covariance matrix in Bayenv2.0, which will result in a reduction in power. However, including a sufficient number of individuals in each pool should mitigate this effect \citep{Zhu2012}. Furthermore, our model assumes perfectly called bases since we do not consider quality scores or sequencing errors. Rearchers dealing with NGS data should exercise caution with these issues. However, examining multiple population pools simultaneously provides some straightforward approaches to minimize error rates in SNP calling, such as calling only SNPs supported by a minimum number of reads in at least one population \citep{Futschik2010}. Such strategies are already good practice in studies of pooled samples and should be used in combination with the Bayenv model. For the application to individual based NGS data, further possible extensions of our model include sequencing errors and probabilistic genotype calling \citep[see][for a discussion on SNP calling from NGS data]{Nielsen2011}.

\paragraph{Outlook}

The population genomic comparison of closely related populations that differ strongly in environmental variables has already yielded many great candidate loci \citep[see for example, altitude adaptation in Tibetans,][]{Beall2010,Simonson2010,Yi2010}. The methods developed here and elsewhere are part of realizing the power of these population comparisons. Such empirical studies also highlight the current deficiencies of such methods, as some of the best signals in these studies are not shared across populations with broadly similar environments, and instead indicate that adaptation has occurred through independent mutations in the same gene or pathway. For example, high altitude adaptation seems to have a different genetic basis in highland Ethiopian and Andean populations \citep[][]{Bigham2010,Scheinfeldt2012}. Methods based on environmental correlations will fail to detect such cases, unless the data are split into the appropriate geographic subsets \citep[e.g.][]{Hancock2011a} on an appropriate geographic scale \citep{Ralph2010}. While shared standing variation will surely be part of the adaptive response across geographically separated instances of similar environments, ideally we would have methods that could cluster signals at the level of the gene or pathway to allow putative cases of parallel adaptation to be identified. The development of such techniques poses an important challenge for future method development.

\section{Acknowledgements}
We thank Gideon Bradburd, Yaniv Brandvain, Fabian Freund, Chuck Langley, Jonathan Pritchard, Peter Ralph, Jeffrey Ross-Ibarra, Karl Schmid, and Alisa Sedghifar for helpful discussions and comments on earlier versions of the manuscript. We also thank Joseph Pickrell for making $X^TX$ available through the HGDP selection browser. TG was supported by the German Federal Ministery for Education and Research (Synbreed, 0315528D), and by a VolkswagenFoundation scholarship (I/84225) affording him to visit UC Davis. GC was supported by a Sloan Foundation fellowship.

\newpage

\bibliographystyle{genetics} 
\bibliography{bayenv2}

\begin{thebibliography}{87}
\expandafter\ifx\csname natexlab\endcsname\relax\def\natexlab#1{#1}\fi

\bibitem[{{\sc Akey} {\em et~al.\/}(2010){\sc Akey}, {\sc Ruhe}, {\sc Akey},
  {\sc Wong}, {\sc Connelly} {\em et~al.\/}}]{Akey2010}
{\sc Akey, J.~M.}, {\sc A.~L. Ruhe}, {\sc D.~T. Akey}, {\sc A.~K. Wong}, {\sc
  C.~F. Connelly}, {\em et~al.\/}, 2010 {Tracking footprints of artificial
  selection in the dog genome.}
\newblock Proceedings of the National Academy of Sciences of the United States
  of America {\bf 107}: 1160--5.

\bibitem[{{\sc Balding}(2003)}]{Balding2003}
{\sc Balding, D.~J.}, 2003 {Likelihood-based inference for genetic correlation
  coefficients}.
\newblock Theoretical Population Biology {\bf 63}: 221--230.

\bibitem[{{\sc Banta} {\em et~al.\/}(2012){\sc Banta}, {\sc Ehrenreich}, {\sc
  Gerard}, {\sc Chou}, {\sc Wilczek} {\em et~al.\/}}]{Banta2012}
{\sc Banta, J.~A.}, {\sc I.~M. Ehrenreich}, {\sc S.~Gerard}, {\sc L.~Chou},
  {\sc A.~Wilczek}, {\em et~al.\/}, 2012 {Climate envelope modelling reveals
  intraspecific relationships among flowering phenology, niche breadth and
  potential range size in {\em Arabidopsis thaliana}.}
\newblock Ecology Letters {\bf 15}: 769--77.

\bibitem[{{\sc Beall} {\em et~al.\/}(2010){\sc Beall}, {\sc Cavalleri}, {\sc
  Deng}, {\sc Elston}, {\sc Gao} {\em et~al.\/}}]{Beall2010}
{\sc Beall, C.~M.}, {\sc G.~L. Cavalleri}, {\sc L.~Deng}, {\sc R.~C. Elston},
  {\sc Y.~Gao}, {\em et~al.\/}, 2010 {Natural selection on {\em EPAS1} ({\em
  HIF2alpha}) associated with low hemoglobin concentration in Tibetan
  highlanders.}
\newblock Proceedings of the National Academy of Sciences of the United States
  of America {\bf 107}: 11459--64.

\bibitem[{{\sc Bhatia} {\em et~al.\/}(2011){\sc Bhatia}, {\sc Patterson}, {\sc
  Pasaniuc}, {\sc Zaitlen}, {\sc Genovese} {\em et~al.\/}}]{Bhatia2011}
{\sc Bhatia, G.}, {\sc N.~Patterson}, {\sc B.~Pasaniuc}, {\sc N.~Zaitlen}, {\sc
  G.~Genovese}, {\em et~al.\/}, 2011 {Genome-wide Comparison of
  African-Ancestry Populations from CARe and Other Cohorts Reveals Signals of
  Natural Selection.}
\newblock American Journal of Human Genetics {\bf 89}: 368--81.

\bibitem[{{\sc Bigham} {\em et~al.\/}(2010){\sc Bigham}, {\sc Bauchet}, {\sc
  Pinto}, {\sc Mao}, {\sc Akey} {\em et~al.\/}}]{Bigham2010}
{\sc Bigham, A.}, {\sc M.~Bauchet}, {\sc D.~Pinto}, {\sc X.~Mao}, {\sc J.~M.
  Akey}, {\em et~al.\/}, 2010 {Identifying Signatures of Natural Selection in
  Tibetan and Andean Populations Using Dense Genome Scan Data}.
\newblock PLoS Genetics {\bf 6}: e1001116.

\bibitem[{{\sc Boitard} {\em et~al.\/}(2012){\sc Boitard}, {\sc
  Schl\"{o}tterer}, {\sc Nolte}, {\sc Pandey} and {\sc Futschik}}]{Boitard2012}
{\sc Boitard, S.}, {\sc C.~Schl\"{o}tterer}, {\sc V.~Nolte}, {\sc R.~Pandey},
  and {\sc A.~Futschik}, 2012 {Detecting selective sweeps from pooled next
  generation sequencing samples.}
\newblock Molecular Biology and Evolution .

\bibitem[{{\sc Bonhomme} {\em et~al.\/}(2010){\sc Bonhomme}, {\sc Chevalet},
  {\sc Servin}, {\sc Boitard}, {\sc Abdallah} {\em et~al.\/}}]{Bonhomme2010}
{\sc Bonhomme, M.}, {\sc C.~Chevalet}, {\sc B.~Servin}, {\sc S.~Boitard}, {\sc
  J.~Abdallah}, {\em et~al.\/}, 2010 {Detecting selection in population trees:
  the Lewontin and Krakauer test extended.}
\newblock Genetics {\bf 186}: 241--262.

\bibitem[{{\sc Boyko} {\em et~al.\/}(2010){\sc Boyko}, {\sc Quignon}, {\sc Li},
  {\sc Schoenebeck}, {\sc Degenhardt} {\em et~al.\/}}]{Boyko2010}
{\sc Boyko, A.~R.}, {\sc P.~Quignon}, {\sc L.~Li}, {\sc J.~J. Schoenebeck},
  {\sc J.~D. Degenhardt}, {\em et~al.\/}, 2010 {A simple genetic architecture
  underlies morphological variation in dogs.}
\newblock PLoS Biology {\bf 8}: e1000451.

\bibitem[{{\sc Cao} {\em et~al.\/}(2011){\sc Cao}, {\sc Schneeberger}, {\sc
  Ossowski}, {\sc G\"{u}nther}, {\sc Bender} {\em et~al.\/}}]{Cao2011}
{\sc Cao, J.}, {\sc K.~Schneeberger}, {\sc S.~Ossowski}, {\sc T.~G\"{u}nther},
  {\sc S.~Bender}, {\em et~al.\/}, 2011 {Whole-genome sequencing of multiple
  {\em Arabidopsis thaliana} populations}.
\newblock Nature Genetics {\bf 43}: 956--963.

\bibitem[{{\sc Cavalli-Sforza}(1966)}]{Cavalli-Sforza1966}
{\sc Cavalli-Sforza, L.~L.}, 1966 {Population structure and human evolution.}
\newblock Proceedings of the Royal Society of London. Series B, Containing
  papers of a Biological character. Royal Society (Great Britain) {\bf 164}:
  362--79.

\bibitem[{{\sc Cheng} {\em et~al.\/}(2012){\sc Cheng}, {\sc White}, {\sc
  Kamdem}, {\sc Mockaitis}, {\sc Costantini} {\em et~al.\/}}]{Cheng2011}
{\sc Cheng, C.}, {\sc B.~J. White}, {\sc C.~Kamdem}, {\sc K.~Mockaitis}, {\sc
  C.~Costantini}, {\em et~al.\/}, 2012 {Ecological Genomics of {\em Anopheles
  gambiae} Along a Latitudinal Cline in Cameroon: A Population Resequencing
  Approach.}
\newblock Genetics {\bf 190}: 1417--1432.

\bibitem[{{\sc Conrad} {\em et~al.\/}(2006){\sc Conrad}, {\sc Jakobsson}, {\sc
  Coop}, {\sc Wen}, {\sc Wall} {\em et~al.\/}}]{Conrad2006}
{\sc Conrad, D.~F.}, {\sc M.~Jakobsson}, {\sc G.~Coop}, {\sc X.~Wen}, {\sc
  J.~D. Wall}, {\em et~al.\/}, 2006 {A worldwide survey of haplotype variation
  and linkage disequilibrium in the human genome.}
\newblock Nature Genetics {\bf 38}: 1251--60.

\bibitem[{{\sc Coop} {\em et~al.\/}(2009){\sc Coop}, {\sc Pickrell}, {\sc
  Novembre}, {\sc Kudaravalli}, {\sc Li} {\em et~al.\/}}]{Coop2009}
{\sc Coop, G.}, {\sc J.~K. Pickrell}, {\sc J.~Novembre}, {\sc S.~Kudaravalli},
  {\sc J.~Li}, {\em et~al.\/}, 2009 {The role of geography in human
  adaptation.}
\newblock PLoS Genetics {\bf 5}: e1000500.

\bibitem[{{\sc Coop} {\em et~al.\/}(2010){\sc Coop}, {\sc Witonsky}, {\sc {Di
  Rienzo}} and {\sc Pritchard}}]{Coop2010}
{\sc Coop, G.}, {\sc D.~Witonsky}, {\sc A.~{Di Rienzo}}, and {\sc J.~K.
  Pritchard}, 2010 {Using environmental correlations to identify loci
  underlying local adaptation.}
\newblock Genetics {\bf 185}: 1411--23.

\bibitem[{{\sc Cutler} and {\sc Jensen}(2010)}]{Cutler2010}
{\sc Cutler, D.~J.}, and {\sc J.~D. Jensen}, 2010 {To pool, or not to pool?}
\newblock Genetics {\bf 186}: 41--3.

\bibitem[{{\sc Eckert} {\em et~al.\/}(2010){\sc Eckert}, {\sc Bower}, {\sc
  Gonz\'{a}lez-Mart\'{\i}nez}, {\sc Wegrzyn}, {\sc Coop} {\em
  et~al.\/}}]{Eckert2010}
{\sc Eckert, A.~J.}, {\sc A.~D. Bower}, {\sc S.~C. Gonz\'{a}lez-Mart\'{\i}nez},
  {\sc J.~L. Wegrzyn}, {\sc G.~Coop}, {\em et~al.\/}, 2010 {Back to nature:
  ecological genomics of loblolly pine ({\em Pinus taeda}, Pinaceae).}
\newblock Molecular Ecology {\bf 19}: 3789--805.

\bibitem[{{\sc Excoffier} {\em et~al.\/}(2009){\sc Excoffier}, {\sc Hofer} and
  {\sc Foll}}]{Excoffier2009}
{\sc Excoffier, L.}, {\sc T.~Hofer}, and {\sc M.~Foll}, 2009 {Detecting loci
  under selection in a hierarchically structured population.}
\newblock Heredity {\bf 103}: 285--98.

\bibitem[{{\sc Fabian} {\em et~al.\/}(2012){\sc Fabian}, {\sc Kapun}, {\sc
  Nolte}, {\sc Kofler}, {\sc Schmidt} {\em et~al.\/}}]{Fabian2012}
{\sc Fabian, D.~K.}, {\sc M.~Kapun}, {\sc V.~Nolte}, {\sc R.~Kofler}, {\sc
  P.~S. Schmidt}, {\em et~al.\/}, 2012 {Genome-wide patterns of latitudinal
  differentiation among populations of {\em Drosophila melanogaster} from North
  America}.
\newblock Molecular Ecology : n/a--n/a.

\bibitem[{{\sc Fang} {\em et~al.\/}(2012){\sc Fang}, {\sc Pyh\"{a}j\"{a}rvi},
  {\sc Weber}, {\sc Dawe}, {\sc Glaubitz} {\em et~al.\/}}]{Fang2012}
{\sc Fang, Z.}, {\sc T.~Pyh\"{a}j\"{a}rvi}, {\sc A.~L. Weber}, {\sc R.~K.
  Dawe}, {\sc J.~C. Glaubitz}, {\em et~al.\/}, 2012 {Megabase-scale Inversion
  Polymorphism in the Wild Ancestor of Maize.}
\newblock Genetics {\bf 191}: 883--894.

\bibitem[{{\sc Flicek} {\em et~al.\/}(2012){\sc Flicek}, {\sc Amode}, {\sc
  Barrell}, {\sc Beal}, {\sc Brent} {\em et~al.\/}}]{Flicek2012}
{\sc Flicek, P.}, {\sc M.~R. Amode}, {\sc D.~Barrell}, {\sc K.~Beal}, {\sc
  S.~Brent}, {\em et~al.\/}, 2012 {Ensembl 2012.}
\newblock Nucleic Acids Research {\bf 40}: D84--90.

\bibitem[{{\sc Fournier-Level} {\em et~al.\/}(2011){\sc Fournier-Level}, {\sc
  Korte}, {\sc Cooper}, {\sc Nordborg}, {\sc Schmitt} {\em
  et~al.\/}}]{Fournier-Level2011}
{\sc Fournier-Level, A.}, {\sc A.~Korte}, {\sc M.~D. Cooper}, {\sc
  M.~Nordborg}, {\sc J.~Schmitt}, {\em et~al.\/}, 2011 {A map of local
  adaptation in {\em Arabidopsis thaliana}.}
\newblock Science {\bf 334}: 86--9.

\bibitem[{{\sc Frichot} {\em et~al.\/}(2012){\sc Frichot}, {\sc Schoville},
  {\sc Bouchard} and {\sc Fran\c{c}ois}}]{Frichot2012}
{\sc Frichot, E.}, {\sc S.~Schoville}, {\sc G.~Bouchard}, and {\sc
  O.~Fran\c{c}ois}, 2012 {Landscape genomic tests for associations between loci
  and environmental gradients} .

\bibitem[{{\sc Fumagalli} {\em et~al.\/}(2011){\sc Fumagalli}, {\sc Sironi},
  {\sc Pozzoli}, {\sc Ferrer-Admettla}, {\sc Pattini} {\em
  et~al.\/}}]{Fumagalli2011}
{\sc Fumagalli, M.}, {\sc M.~Sironi}, {\sc U.~Pozzoli}, {\sc
  A.~Ferrer-Admettla}, {\sc L.~Pattini}, {\em et~al.\/}, 2011 {Signatures of
  Environmental Genetic Adaptation Pinpoint Pathogens as the Main Selective
  Pressure through Human Evolution}.
\newblock PLoS Genetics {\bf 7}: e1002355.

\bibitem[{{\sc Futschik} and {\sc Schl\"{o}tterer}(2010)}]{Futschik2010}
{\sc Futschik, A.}, and {\sc C.~Schl\"{o}tterer}, 2010 {The next generation of
  molecular markers from massively parallel sequencing of pooled DNA samples.}
\newblock Genetics {\bf 186}: 207--18.

\bibitem[{{\sc Gasper} and {\sc Swanson}(2006)}]{Gasper2006}
{\sc Gasper, J.}, and {\sc W.~J. Swanson}, 2006 {Molecular population genetics
  of the gene encoding the human fertilization protein zonadhesin reveals rapid
  adaptive evolution.}
\newblock American Journal of Human Genetics {\bf 79}: 820--30.

\bibitem[{{\sc Guillot}(2012)}]{Guillot2012}
{\sc Guillot, G.}, 2012 {Detection of correlation between genotypes and
  environmental variables. A fast computational approach for genomewide
  studies} .

\bibitem[{{\sc Guillot} and {\sc Rousset}(2011)}]{Guillot2011}
{\sc Guillot, G.}, and {\sc F.~Rousset}, 2011 {On the use of the simple and
  partial Mantel tests in presence of spatial auto-correlation} .

\bibitem[{{\sc Hancock} {\em et~al.\/}(2011{\natexlab{a}}){\sc Hancock}, {\sc
  Brachi}, {\sc Faure}, {\sc Horton}, {\sc Jarymowycz} {\em
  et~al.\/}}]{Hancock2011}
{\sc Hancock, A.~M.}, {\sc B.~Brachi}, {\sc N.~Faure}, {\sc M.~W. Horton}, {\sc
  L.~B. Jarymowycz}, {\em et~al.\/}, 2011{\natexlab{a}} {Adaptation to Climate
  Across the {\em Arabidopsis thaliana} Genome}.
\newblock Science {\bf 334}: 83--86.

\bibitem[{{\sc Hancock} {\em et~al.\/}(2011{\natexlab{b}}){\sc Hancock}, {\sc
  Clark}, {\sc Qian} and {\sc {Di Rienzo}}}]{Hancock2011b}
{\sc Hancock, A.~M.}, {\sc V.~J. Clark}, {\sc Y.~Qian}, and {\sc A.~{Di
  Rienzo}}, 2011{\natexlab{b}} {Population genetic analysis of the uncoupling
  proteins supports a role for {\em UCP3} in human cold resistance.}
\newblock Molecular Biology and Evolution {\bf 28}: 601--14.

\bibitem[{{\sc Hancock} {\em et~al.\/}(2011{\natexlab{c}}){\sc Hancock}, {\sc
  Witonsky}, {\sc Alkorta-Aranburu}, {\sc Beall}, {\sc Gebremedhin} {\em
  et~al.\/}}]{Hancock2011a}
{\sc Hancock, A.~M.}, {\sc D.~B. Witonsky}, {\sc G.~Alkorta-Aranburu}, {\sc
  C.~M. Beall}, {\sc A.~Gebremedhin}, {\em et~al.\/}, 2011{\natexlab{c}}
  {Adaptations to Climate-Mediated Selective Pressures in Humans}.
\newblock PLoS Genetics {\bf 7}: e1001375.

\bibitem[{{\sc Hancock} {\em et~al.\/}(2010){\sc Hancock}, {\sc Witonsky}, {\sc
  Ehler}, {\sc Alkorta-Aranburu}, {\sc Beall} {\em et~al.\/}}]{Hancock2010}
{\sc Hancock, A.~M.}, {\sc D.~B. Witonsky}, {\sc E.~Ehler}, {\sc
  G.~Alkorta-Aranburu}, {\sc C.~Beall}, {\em et~al.\/}, 2010 {Human adaptations
  to diet, subsistence, and ecoregion are due to subtle shifts in allele
  frequency.}
\newblock Proceedings of the National Academy of Sciences of the United States
  of America {\bf 107 Suppl}: 8924--30.

\bibitem[{{\sc Hancock} {\em et~al.\/}(2008){\sc Hancock}, {\sc Witonsky}, {\sc
  Gordon}, {\sc Eshel}, {\sc Pritchard} {\em et~al.\/}}]{Hancock2008}
{\sc Hancock, A.~M.}, {\sc D.~B. Witonsky}, {\sc A.~S. Gordon}, {\sc G.~Eshel},
  {\sc J.~K. Pritchard}, {\em et~al.\/}, 2008 {Adaptations to climate in
  candidate genes for common metabolic disorders.}
\newblock PLoS Genetics {\bf 4}: e32.

\bibitem[{{\sc He} {\em et~al.\/}(2011){\sc He}, {\sc Zhai}, {\sc Wen}, {\sc
  Tang}, {\sc Wang} {\em et~al.\/}}]{He2011}
{\sc He, Z.}, {\sc W.~Zhai}, {\sc H.~Wen}, {\sc T.~Tang}, {\sc Y.~Wang}, {\em
  et~al.\/}, 2011 {Two Evolutionary Histories in the Genome of Rice: the Roles
  of Domestication Genes}.
\newblock PLoS Genetics {\bf 7}: e1002100.

\bibitem[{{\sc Hernandez} {\em et~al.\/}(2011){\sc Hernandez}, {\sc Kelley},
  {\sc Elyashiv}, {\sc Melton}, {\sc Auton} {\em et~al.\/}}]{Hernandez2011}
{\sc Hernandez, R.~D.}, {\sc J.~L. Kelley}, {\sc E.~Elyashiv}, {\sc S.~C.
  Melton}, {\sc A.~Auton}, {\em et~al.\/}, 2011 {Classic selective sweeps were
  rare in recent human evolution.}
\newblock Science {\bf 331}: 920--4.

\bibitem[{{\sc Hinohara} {\em et~al.\/}(2009){\sc Hinohara}, {\sc Nakajima},
  {\sc Yasunami}, {\sc Houda}, {\sc Sasaoka} {\em et~al.\/}}]{Hinohara2009}
{\sc Hinohara, K.}, {\sc T.~Nakajima}, {\sc M.~Yasunami}, {\sc S.~Houda}, {\sc
  T.~Sasaoka}, {\em et~al.\/}, 2009 {Megakaryoblastic leukemia factor-1 gene in
  the susceptibility to coronary artery disease.}
\newblock Human Genetics {\bf 126}: 539--47.

\bibitem[{{\sc Hoffmann} and {\sc Weeks}(2007)}]{Hoffmann2007}
{\sc Hoffmann, A.~A.}, and {\sc A.~R. Weeks}, 2007 {Climatic selection on genes
  and traits after a 100 year-old invasion: a critical look at the
  temperate-tropical clines in {\em Drosophila melanogaster} from eastern
  Australia.}
\newblock Genetica {\bf 129}: 133--47.

\bibitem[{{\sc Huxley}(1939)}]{Huxley1939}
{\sc Huxley, J.~S.}, 1939 {Clines: an auxiliary method in taxonomy}.
\newblock Bijdr. Dierk {\bf 27}: 491----520.

\bibitem[{{\sc Jablonski}(2004)}]{Jablonski2004}
{\sc Jablonski, N.~G.}, 2004 {the Evolution of Human Skin and Skin Color}.
\newblock Annual Review of Anthropology {\bf 33}: 585--623.

\bibitem[{{\sc Jones} {\em et~al.\/}(2011){\sc Jones}, {\sc Chan}, {\sc
  Schmutz}, {\sc Grimwood}, {\sc Brady} {\em et~al.\/}}]{Jones2011}
{\sc Jones, F.~C.}, {\sc Y.~F. Chan}, {\sc J.~Schmutz}, {\sc J.~Grimwood}, {\sc
  S.~D. Brady}, {\em et~al.\/}, 2011 {A Genome-wide SNP Genotyping Array
  Reveals Patterns of Global and Repeated Species-Pair Divergence in
  Sticklebacks.}
\newblock Current Biology {\bf 22}: 83--90.

\bibitem[{{\sc Kang} {\em et~al.\/}(2010){\sc Kang}, {\sc Sul}, {\sc Service},
  {\sc Zaitlen}, {\sc Kong} {\em et~al.\/}}]{Kang2010}
{\sc Kang, H.~M.}, {\sc J.~H. Sul}, {\sc S.~K. Service}, {\sc N.~A. Zaitlen},
  {\sc S.-Y. Kong}, {\em et~al.\/}, 2010 {Variance component model to account
  for sample structure in genome-wide association studies.}
\newblock Nature Genetics {\bf 42}: 348--54.

\bibitem[{{\sc Kang} {\em et~al.\/}(2008){\sc Kang}, {\sc Zaitlen}, {\sc Wade},
  {\sc Kirby}, {\sc Heckerman} {\em et~al.\/}}]{Kang2008}
{\sc Kang, H.~M.}, {\sc N.~a. Zaitlen}, {\sc C.~M. Wade}, {\sc A.~Kirby}, {\sc
  D.~Heckerman}, {\em et~al.\/}, 2008 {Efficient control of population
  structure in model organism association mapping.}
\newblock Genetics {\bf 178}: 1709--23.

\bibitem[{{\sc Keller} {\em et~al.\/}(2012){\sc Keller}, {\sc Levsen}, {\sc
  Olson} and {\sc Tiffin}}]{Keller2012}
{\sc Keller, S.~R.}, {\sc N.~Levsen}, {\sc M.~S. Olson}, and {\sc P.~Tiffin},
  2012 {Local Adaptation in the Flowering-Time Gene Network of Balsam Poplar,
  {\em Populus balsamifera L.}}
\newblock Molecular Biology and evolution .

\bibitem[{{\sc Kofler} {\em et~al.\/}(2012){\sc Kofler}, {\sc Betancourt} and
  {\sc Schl\"{o}tterer}}]{Kofler2012}
{\sc Kofler, R.}, {\sc A.~J. Betancourt}, and {\sc C.~Schl\"{o}tterer}, 2012
  {Sequencing of Pooled DNA Samples (Pool-Seq) Uncovers Complex Dynamics of
  Transposable Element Insertions in {\em Drosophila melanogaster}}.
\newblock PLoS Genetics {\bf 8}: e1002487.

\bibitem[{{\sc Kolaczkowski} {\em et~al.\/}(2011){\sc Kolaczkowski}, {\sc
  Kern}, {\sc Holloway} and {\sc Begun}}]{Kolaczkowski2011}
{\sc Kolaczkowski, B.}, {\sc A.~D. Kern}, {\sc A.~K. Holloway}, and {\sc D.~J.
  Begun}, 2011 {Genomic differentiation between temperate and tropical
  Australian populations of {\em Drosophila melanogaster}.}
\newblock Genetics {\bf 187}: 245--60.

\bibitem[{{\sc Kring} {\em et~al.\/}(2008){\sc Kring}, {\sc Larsen}, {\sc
  Holst}, {\sc Toubro}, {\sc Hansen} {\em et~al.\/}}]{Kring2008}
{\sc Kring, S. I.~I.}, {\sc L.~H. Larsen}, {\sc C.~Holst}, {\sc S.~r. Toubro},
  {\sc T.~Hansen}, {\em et~al.\/}, 2008 {Genotype-phenotype associations in
  obesity dependent on definition of the obesity phenotype.}
\newblock Obesity Facts {\bf 1}: 138--45.

\bibitem[{{\sc Lamason} {\em et~al.\/}(2005){\sc Lamason}, {\sc Mohideen}, {\sc
  Mest}, {\sc Wong}, {\sc Norton} {\em et~al.\/}}]{Lamason2005}
{\sc Lamason, R.~L.}, {\sc M.-A. P.~K. Mohideen}, {\sc J.~R. Mest}, {\sc A.~C.
  Wong}, {\sc H.~L. Norton}, {\em et~al.\/}, 2005 {{\em SLC24A5}, a putative
  cation exchanger, affects pigmentation in zebrafish and humans.}
\newblock Science {\bf 310}: 1782--6.

\bibitem[{{\sc Lewontin} and {\sc Krakauer}(1973)}]{Lewontin1973}
{\sc Lewontin, R.~C.}, and {\sc J.~Krakauer}, 1973 {Distribution of gene
  frequency as a test of the theory of the selective neutrality of
  polymorphisms.}
\newblock Genetics {\bf 74}: 175--95.

\bibitem[{{\sc Li} {\em et~al.\/}(2008){\sc Li}, {\sc Absher}, {\sc Tang}, {\sc
  Southwick}, {\sc Casto} {\em et~al.\/}}]{Li2008}
{\sc Li, J.~Z.}, {\sc D.~M. Absher}, {\sc H.~Tang}, {\sc A.~M. Southwick}, {\sc
  A.~M. Casto}, {\em et~al.\/}, 2008 {Worldwide human relationships inferred
  from genome-wide patterns of variation.}
\newblock Science {\bf 319}: 1100--4.

\bibitem[{{\sc Limborg} {\em et~al.\/}(2012){\sc Limborg}, {\sc Blankenship},
  {\sc Young}, {\sc Utter}, {\sc Seeb} {\em et~al.\/}}]{Limborg2012}
{\sc Limborg, M.~T.}, {\sc S.~M. Blankenship}, {\sc S.~F. Young}, {\sc F.~M.
  Utter}, {\sc L.~W. Seeb}, {\em et~al.\/}, 2012 {Signatures of natural
  selection among lineages and habitats in {\em Oncorhynchus mykiss}.}
\newblock Ecology and Evolution {\bf 2}: 1--18.

\bibitem[{{\sc Lindgren} {\em et~al.\/}(2011){\sc Lindgren}, {\sc Rue} and {\sc
  Lindstr\"{o}m}}]{Lindgren2011}
{\sc Lindgren, F.}, {\sc H.~v. Rue}, and {\sc J.~Lindstr\"{o}m}, 2011 {An
  explicit link between Gaussian fields and Gaussian Markov random fields: the
  stochastic partial differential equation approach}.
\newblock Journal of the Royal Statistical Society: Series B (Statistical
  Methodology) {\bf 73}: 423--498.

\bibitem[{{\sc Long} {\em et~al.\/}(2011){\sc Long}, {\sc Jeffares}, {\sc
  Zhang}, {\sc Ye}, {\sc Nizhynska} {\em et~al.\/}}]{Long2011}
{\sc Long, Q.}, {\sc D.~C. Jeffares}, {\sc Q.~Zhang}, {\sc K.~Ye}, {\sc
  V.~Nizhynska}, {\em et~al.\/}, 2011 {PoolHap: Inferring Haplotype Frequencies
  from Pooled Samples by Next Generation Sequencing}.
\newblock PLoS ONE {\bf 6}: e15292.

\bibitem[{{\sc Ma} {\em et~al.\/}(2001){\sc Ma}, {\sc Morris}, {\sc Valentine},
  {\sc Li}, {\sc Herbrick} {\em et~al.\/}}]{Ma2001}
{\sc Ma, Z.}, {\sc S.~W. Morris}, {\sc V.~Valentine}, {\sc M.~Li}, {\sc J.~A.
  Herbrick}, {\em et~al.\/}, 2001 {Fusion of two novel genes, {\em RBM15} and
  {\em MKL1}, in the t(1;22)(p13;q13) of acute megakaryoblastic leukemia.}
\newblock Nature Genetics {\bf 28}: 220--1.

\bibitem[{{\sc Maglott} {\em et~al.\/}(2011){\sc Maglott}, {\sc Ostell}, {\sc
  Pruitt} and {\sc Tatusova}}]{Maglott2011}
{\sc Maglott, D.}, {\sc J.~Ostell}, {\sc K.~D. Pruitt}, and {\sc T.~Tatusova},
  2011 {Entrez Gene: gene-centered information at NCBI.}
\newblock Nucleic Acids Research {\bf 39}: D52--7.

\bibitem[{{\sc Marsh} {\em et~al.\/}(2002){\sc Marsh}, {\sc Weingarth}, {\sc
  Novi}, {\sc Chen}, {\sc Trumbauer} {\em et~al.\/}}]{Marsh2002}
{\sc Marsh, D.~J.}, {\sc D.~T. Weingarth}, {\sc D.~E. Novi}, {\sc H.~Y. Chen},
  {\sc M.~E. Trumbauer}, {\em et~al.\/}, 2002 {Melanin-concentrating hormone 1
  receptor-deficient mice are lean, hyperactive, and hyperphagic and have
  altered metabolism.}
\newblock Proceedings of the National Academy of Sciences of the United States
  of America {\bf 99}: 3240--5.

\bibitem[{{\sc Nakayama} {\em et~al.\/}(2002){\sc Nakayama}, {\sc Fukamachi},
  {\sc Kimura}, {\sc Koda}, {\sc Soemantri} {\em et~al.\/}}]{Nakayama2002}
{\sc Nakayama, K.}, {\sc S.~Fukamachi}, {\sc H.~Kimura}, {\sc Y.~Koda}, {\sc
  A.~Soemantri}, {\em et~al.\/}, 2002 {Distinctive distribution of AIM1
  polymorphism among major human populations with different skin color.}
\newblock Journal of Human Genetics {\bf 47}: 92--4.

\bibitem[{{\sc Nicholson} {\em et~al.\/}(2002){\sc Nicholson}, {\sc Smith},
  {\sc Jonsson}, {\sc Gustafsson}, {\sc Stefansson} {\em
  et~al.\/}}]{Nicholson2002}
{\sc Nicholson, G.}, {\sc A.~V. Smith}, {\sc F.~Jonsson}, {\sc O.~Gustafsson},
  {\sc K.~Stefansson}, {\em et~al.\/}, 2002 {Assessing population
  differentiation and isolation from single-nucleotide polymorphism data}.
\newblock Journal of the Royal Statistical Society: Series B {\bf 64}:
  695--715.

\bibitem[{{\sc Nielsen} {\em et~al.\/}(2011){\sc Nielsen}, {\sc Paul}, {\sc
  Albrechtsen} and {\sc Song}}]{Nielsen2011}
{\sc Nielsen, R.}, {\sc J.~S. Paul}, {\sc A.~Albrechtsen}, and {\sc Y.~S.
  Song}, 2011 {Genotype and SNP calling from next-generation sequencing data}.
\newblock Nature Reviews Genetics {\bf 12}: 443--451.

\bibitem[{{\sc Ohashi} {\em et~al.\/}(2011){\sc Ohashi}, {\sc Naka} and {\sc
  Tsuchiya}}]{Ohashi2011}
{\sc Ohashi, J.}, {\sc I.~Naka}, and {\sc N.~Tsuchiya}, 2011 {The impact of
  natural selection on an {\em ABCC11} SNP determining earwax type.}
\newblock Molecular Biology and Evolution {\bf 28}: 849--57.

\bibitem[{{\sc Orozco-Terwengel} {\em et~al.\/}(2012){\sc Orozco-Terwengel},
  {\sc Kapun}, {\sc Nolte}, {\sc Kofler}, {\sc Flatt} {\em
  et~al.\/}}]{Orozco-Terwengel2012}
{\sc Orozco-Terwengel, P.}, {\sc M.~Kapun}, {\sc V.~Nolte}, {\sc R.~Kofler},
  {\sc T.~Flatt}, {\em et~al.\/}, 2012 {Adaptation of {\em Drosophila} to a
  novel laboratory environment reveals temporally heterogeneous trajectories of
  selected alleles.}
\newblock Molecular Ecology .

\bibitem[{{\sc Parmacek}(2007)}]{Parmacek2007}
{\sc Parmacek, M.~S.}, 2007 {Myocardin-related transcription factors: critical
  coactivators regulating cardiovascular development and adaptation.}
\newblock Circulation Research {\bf 100}: 633--44.

\bibitem[{{\sc Pickrell} {\em et~al.\/}(2009){\sc Pickrell}, {\sc Coop}, {\sc
  Novembre}, {\sc Kudaravalli}, {\sc Li} {\em et~al.\/}}]{Pickrell2009}
{\sc Pickrell, J.~K.}, {\sc G.~Coop}, {\sc J.~Novembre}, {\sc S.~Kudaravalli},
  {\sc J.~Z. Li}, {\em et~al.\/}, 2009 {Signals of recent positive selection in
  a worldwide sample of human populations.}
\newblock Genome Research {\bf 19}: 826--37.

\bibitem[{{\sc Pickrell} and {\sc Pritchard}(2012)}]{Pickrell2012}
{\sc Pickrell, J.~K.}, and {\sc J.~K. Pritchard}, 2012 {Inference of population
  splits and mixtures from genome-wide allele frequency data} .

\bibitem[{{\sc Price} {\em et~al.\/}(2009){\sc Price}, {\sc Helgason}, {\sc
  Palsson}, {\sc Stefansson}, {\sc {St Clair}} {\em et~al.\/}}]{Price2009}
{\sc Price, A.~L.}, {\sc A.~Helgason}, {\sc S.~Palsson}, {\sc H.~Stefansson},
  {\sc D.~{St Clair}}, {\em et~al.\/}, 2009 {The impact of divergence time on
  the nature of population structure: an example from Iceland.}
\newblock PLoS Genetics {\bf 5}: e1000505.

\bibitem[{{\sc Pritchard} {\em et~al.\/}(2010){\sc Pritchard}, {\sc Pickrell}
  and {\sc Coop}}]{Pritchard2010}
{\sc Pritchard, J.~K.}, {\sc J.~K. Pickrell}, and {\sc G.~Coop}, 2010 {The
  genetics of human adaptation: hard sweeps, soft sweeps, and polygenic
  adaptation.}
\newblock Current Biology {\bf 20}: R208--15.

\bibitem[{{\sc Pyh\"{a}j\"{a}rvi} {\em et~al.\/}(2012){\sc Pyh\"{a}j\"{a}rvi},
  {\sc Hufford}, {\sc Mezmouk} and {\sc Ross-Ibarra}}]{Pyhajarvi2012}
{\sc Pyh\"{a}j\"{a}rvi, T.}, {\sc M.~B. Hufford}, {\sc S.~Mezmouk}, and {\sc
  J.~Ross-Ibarra}, 2012 {Complex patterns of local adaptation in teosinte} .

\bibitem[{{\sc {R Development Core Team}}(2011)}]{RDevelopmentCoreTeam2009}
{\sc {R Development Core Team}}, 2011 {\em {R: A Language and Environment for
  Statistical Computing}\/}.
\newblock R Foundation for Statistical Computing, Vienna, Austria.

\bibitem[{{\sc Ralph} and {\sc Coop}(2010)}]{Ralph2010}
{\sc Ralph, P.}, and {\sc G.~Coop}, 2010 {Parallel adaptation: one or many
  waves of advance of an advantageous allele?}
\newblock Genetics {\bf 186}: 647--68.

\bibitem[{{\sc Robertson}(1975)}]{Robertson1975}
{\sc Robertson, A.}, 1975 {Gene frequency distributions as a test of selective
  neutrality.}
\newblock Genetics {\bf 81}: 775--785.

\bibitem[{{\sc Rosenberg} {\em et~al.\/}(2002){\sc Rosenberg}, {\sc Pritchard},
  {\sc Weber}, {\sc Cann}, {\sc Kidd} {\em et~al.\/}}]{Rosenberg2002}
{\sc Rosenberg, N.~A.}, {\sc J.~K. Pritchard}, {\sc J.~L. Weber}, {\sc H.~M.
  Cann}, {\sc K.~K. Kidd}, {\em et~al.\/}, 2002 {Genetic structure of human
  populations.}
\newblock Science {\bf 298}: 2381--5.

\bibitem[{{\sc Rue} {\em et~al.\/}(2009){\sc Rue}, {\sc Martino} and {\sc
  Chopin}}]{Rue2009}
{\sc Rue, H.~v.}, {\sc S.~Martino}, and {\sc N.~Chopin}, 2009 {Approximate
  Bayesian inference for latent Gaussian models by using integrated nested
  Laplace approximations}.
\newblock Journal of the Royal Statistical Society: Series B (Statistical
  Methodology) {\bf 71}: 319--392.

\bibitem[{{\sc Rutanen} {\em et~al.\/}(2007){\sc Rutanen}, {\sc
  Pihlajam\"{a}ki}, {\sc V\"{a}nttinen}, {\sc Salmenniemi}, {\sc Ruotsalainen}
  {\em et~al.\/}}]{Rutanen2007}
{\sc Rutanen, J.}, {\sc J.~Pihlajam\"{a}ki}, {\sc M.~V\"{a}nttinen}, {\sc
  U.~Salmenniemi}, {\sc E.~Ruotsalainen}, {\em et~al.\/}, 2007 {Single
  nucleotide polymorphisms of the {\em MCHR1} gene do not affect metabolism in
  humans.}
\newblock Obesity {\bf 15}: 2902--7.

\bibitem[{{\sc Sabeti} {\em et~al.\/}(2007){\sc Sabeti}, {\sc Varilly}, {\sc
  Fry}, {\sc Lohmueller}, {\sc Hostetter} {\em et~al.\/}}]{Sabeti2007}
{\sc Sabeti, P.~C.}, {\sc P.~Varilly}, {\sc B.~Fry}, {\sc J.~Lohmueller}, {\sc
  E.~Hostetter}, {\em et~al.\/}, 2007 {Genome-wide detection and
  characterization of positive selection in human populations.}
\newblock Nature {\bf 449}: 913--918.

\bibitem[{{\sc Samanta} {\em et~al.\/}(2009){\sc Samanta}, {\sc Li} and {\sc
  Weir}}]{Samanta2009}
{\sc Samanta, S.}, {\sc Y.-J. Li}, and {\sc B.~S. Weir}, 2009 {Drawing
  inferences about the coancestry coefficient.}
\newblock Theoretical Population Biology {\bf 75}: 312--9.

\bibitem[{{\sc Scharenberg} {\em et~al.\/}(2010){\sc Scharenberg}, {\sc
  Chiquet-Ehrismann} and {\sc Asparuhova}}]{Scharenberg2010}
{\sc Scharenberg, M.~A.}, {\sc R.~Chiquet-Ehrismann}, and {\sc M.~B.
  Asparuhova}, 2010 {Megakaryoblastic leukemia protein-1 ({\em MKL1}):
  Increasing evidence for an involvement in cancer progression and metastasis.}
\newblock The International Journal of Biochemistry \& Cell Biology {\bf 42}:
  1911--4.

\bibitem[{{\sc Scheinfeldt} {\em et~al.\/}(2012){\sc Scheinfeldt}, {\sc Soi},
  {\sc Thompson}, {\sc Ranciaro}, {\sc Woldemeskel} {\em
  et~al.\/}}]{Scheinfeldt2012}
{\sc Scheinfeldt, L.~B.}, {\sc S.~Soi}, {\sc S.~Thompson}, {\sc A.~Ranciaro},
  {\sc D.~Woldemeskel}, {\em et~al.\/}, 2012 {Genetic adaptation to high
  altitude in the Ethiopian highlands.}
\newblock Genome Biology {\bf 13}: R1.

\bibitem[{{\sc Simonson} {\em et~al.\/}(2010){\sc Simonson}, {\sc Yang}, {\sc
  Huff}, {\sc Yun}, {\sc Qin} {\em et~al.\/}}]{Simonson2010}
{\sc Simonson, T.~S.}, {\sc Y.~Yang}, {\sc C.~D. Huff}, {\sc H.~Yun}, {\sc
  G.~Qin}, {\em et~al.\/}, 2010 {Genetic evidence for high-altitude adaptation
  in Tibet.}
\newblock Science {\bf 329}: 72--5.

\bibitem[{{\sc Speliotes} {\em et~al.\/}(2010){\sc Speliotes}, {\sc Willer},
  {\sc Berndt}, {\sc Monda}, {\sc Thorleifsson} {\em et~al.\/}}]{Speliotes2010}
{\sc Speliotes, E.~K.}, {\sc C.~J. Willer}, {\sc S.~I. Berndt}, {\sc K.~L.
  Monda}, {\sc G.~Thorleifsson}, {\em et~al.\/}, 2010 {Association analyses of
  249,796 individuals reveal 18 new loci associated with body mass index.}
\newblock Nature Genetics {\bf 42}: 937--48.

\bibitem[{{\sc Stinchcombe} {\em et~al.\/}(2004){\sc Stinchcombe}, {\sc
  Weinig}, {\sc Ungerer}, {\sc Olsen}, {\sc Mays} {\em
  et~al.\/}}]{Stinchcombe2004}
{\sc Stinchcombe, J.~R.}, {\sc C.~Weinig}, {\sc M.~Ungerer}, {\sc K.~M. Olsen},
  {\sc C.~Mays}, {\em et~al.\/}, 2004 {A latitudinal cline in flowering time in
  {\em Arabidopsis thaliana} modulated by the flowering time gene {\em
  FRIGIDA}.}
\newblock Proceedings of the National Academy of Sciences of the United States
  of America {\bf 101}: 4712--7.

\bibitem[{{\sc Turner} {\em et~al.\/}(2010){\sc Turner}, {\sc Bourne}, {\sc
  {Von Wettberg}}, {\sc Hu} and {\sc Nuzhdin}}]{Turner2010}
{\sc Turner, T.~L.}, {\sc E.~C. Bourne}, {\sc E.~J. {Von Wettberg}}, {\sc T.~T.
  Hu}, and {\sc S.~V. Nuzhdin}, 2010 {Population resequencing reveals local
  adaptation of {\em Arabidopsis lyrata} to serpentine soils.}
\newblock Nature Genetics {\bf 42}: 260--3.

\bibitem[{{\sc Turner} {\em et~al.\/}(2011){\sc Turner}, {\sc Stewart}, {\sc
  Fields}, {\sc Rice} and {\sc Tarone}}]{Turner2011}
{\sc Turner, T.~L.}, {\sc A.~D. Stewart}, {\sc A.~T. Fields}, {\sc W.~R. Rice},
  and {\sc A.~M. Tarone}, 2011 {Population-based resequencing of experimentally
  evolved populations reveals the genetic basis of body size variation in {\em
  Drosophila melanogaster}.}
\newblock PLoS Genetics {\bf 7}: e1001336.

\bibitem[{{\sc Weir} and {\sc Hill}(2002)}]{Weir2002}
{\sc Weir, B.~S.}, and {\sc W.~G. Hill}, 2002 {Estimating F-statistics.}
\newblock Annual Review of Genetics {\bf 36}: 721--50.

\bibitem[{{\sc Wermter} {\em et~al.\/}(2005){\sc Wermter}, {\sc Reichwald},
  {\sc B\"{u}ch}, {\sc Geller}, {\sc Platzer} {\em et~al.\/}}]{Wermter2005}
{\sc Wermter, A.-K.}, {\sc K.~Reichwald}, {\sc T.~B\"{u}ch}, {\sc F.~Geller},
  {\sc C.~Platzer}, {\em et~al.\/}, 2005 {Mutation analysis of the {\em MCHR1}
  gene in human obesity.}
\newblock European Journal of Endocrinology {\bf 152}: 851--62.

\bibitem[{{\sc Yi} {\em et~al.\/}(2010){\sc Yi}, {\sc Liang}, {\sc
  Huerta-Sanchez}, {\sc Jin}, {\sc Cuo} {\em et~al.\/}}]{Yi2010}
{\sc Yi, X.}, {\sc Y.~Liang}, {\sc E.~Huerta-Sanchez}, {\sc X.~Jin}, {\sc
  Z.~X.~P. Cuo}, {\em et~al.\/}, 2010 {Sequencing of 50 human exomes reveals
  adaptation to high altitude.}
\newblock Science {\bf 329}: 75--8.

\bibitem[{{\sc Yu} {\em et~al.\/}(2006){\sc Yu}, {\sc Pressoir}, {\sc Briggs},
  {\sc Bi}, {\sc Yamasaki} {\em et~al.\/}}]{Yu2006}
{\sc Yu, J.}, {\sc G.~Pressoir}, {\sc W.~H. Briggs}, {\sc I.~V. Bi}, {\sc
  M.~Yamasaki}, {\em et~al.\/}, 2006 {A unified mixed-model method for
  association mapping that accounts for multiple levels of relatedness}.
\newblock Nature Genetics {\bf 38}: 203--208.

\bibitem[{{\sc Zhou} and {\sc Stephens}(2012)}]{Zhou2012}
{\sc Zhou, X.}, and {\sc M.~Stephens}, 2012 {Genome-wide efficient mixed-model
  analysis for association studies.}
\newblock Nature Genetics {\bf 44}: 821--4.

\bibitem[{{\sc Zhu} {\em et~al.\/}(2012){\sc Zhu}, {\sc Bergland}, {\sc
  Gonz\'{a}lez} and {\sc Petrov}}]{Zhu2012}
{\sc Zhu, Y.}, {\sc A.~O. Bergland}, {\sc J.~Gonz\'{a}lez}, and {\sc D.~A.
  Petrov}, 2012 {Empirical Validation of Pooled Whole Genome Population
  Re-Sequencing in {\em Drosophila melanogaster}}.
\newblock PLoS ONE {\bf 7}: e41901.

\end{thebibliography}

\end{document}